\PassOptionsToPackage{dvipsnames}{xcolor}
\documentclass[sigplan,nonacm,10pt,pbalance]{acmart}
\setcopyright{none}
\AtBeginDocument{
  }

\usepackage{subcaption}
\usepackage{amsmath}
\usepackage{float}
\usepackage{macros}
\renewcommand{\tinyskip}{}
\usepackage[linesnumbered,ruled,vlined]{algorithm2e}
\usepackage{booktabs}
\usepackage{colortbl}
\usepackage{tabularx}
\usepackage{multirow}
\usepackage{booktabs}
\usepackage{tabularx}
\usepackage{booktabs}
\usepackage{array}

\usepackage{xcolor}
\usepackage{eso-pic}

\newcolumntype{Y}{>{\raggedright\arraybackslash}X}
\newcolumntype{P}[1]{>{\raggedright\arraybackslash}p{#1}}

\newcommand{\sys}{\textsc{Nereus}\xspace}

\newcommand{\abs}{{\small \texttt{EMU}}\xspace}
\newcommand{\abss}{{\small \texttt{EMU}s}\xspace}

\fancypagestyle{paperheader}{%
  \fancyhf{}
  \fancyhead[L]{\scriptsize Nereus: Adaptive Parallelism for LLM Post-Training}
  \fancyhead[R]{\footnotesize Songlin Jiang et al.}
  \fancyfoot[C]{\raisebox{-15pt}{\footnotesize\thepage}}
  
}
\fancypagestyle{paperfirst}{%
  \fancyhf{}
  \fancyfoot[C]{\raisebox{-15pt}{\footnotesize\thepage}}
  
}
\newlength{\papercolwidth}
\newlength{\papertextleft}
\newlength{\papercolx}
\AddToShipoutPictureFG{%
  \ifnum\value{page}=1\relax
    \AtPageLowerLeft{%
      \put(\LenToUnit{\papertextleft},\LenToUnit{160pt}){%
        \parbox[t]{\papercolwidth}{%
          \rule{\papercolwidth}{0.4pt}\par\vspace{2.6pt}%
          \includegraphics[height=5ex]{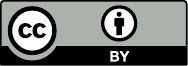}\par
          \footnotesize This work is licensed under a Creative Commons Attribution International 4.0 License.\par
          \textcopyright\ 2026 Copyright held by the owner/author(s).%
        }%
      }%
      \put(\LenToUnit{\papercolx},\LenToUnit{100pt}){%
        \parbox[t]{\papercolwidth}{%
          \vspace{\baselineskip}%
          \rule{4pc}{0.4pt}\par\vspace{2.6pt}%
          \footnotesize *Equal contribution.\par
          \textdagger Corresponding author: bo.zhao@aalto.fi.%
        }%
      }%
    }%
  \fi
}

\makeatletter
\newcommand{\silentcity}{\global\@ACM@citypresenttrue}
\newcommand{\silentcountry}{\global\@ACM@countrypresenttrue}
\makeatother

\begin{document}

\title[]{Nereus: Adaptive Parallelism for LLM Post-Training}

\author{Songlin Jiang\textsuperscript{1,*}\quad Tuo Shi\textsuperscript{2,*}\quad Sitong Zhang\textsuperscript{1}\quad Zeke Wang\textsuperscript{3}\\
Mario Di Francesco\textsuperscript{1}\quad Bo Zhao\textsuperscript{1,\textdagger}}
\affiliation{
  \institution{\textsuperscript{1}Aalto University\quad \textsuperscript{2}Shenzhen University of Advanced Technology\quad \textsuperscript{3}Zhejiang University}
  \silentcity\silentcountry}

\begin{abstract}
Reinforcement learning (RL) post-training for large language models (LLMs) coordinates multiple models across generation, inference,
and training on GPU clusters. Several factors may change during a run, including resource availability, sequence length, memory pressure, and stage bottlenecks. As a consequence, an execution plan that was initially suitable can then become slow or even infeasible over time.
However, adapting a job whose models share GPUs
entails significant challenges: deciding whether a new plan
is worth the transition cost, reusing the job's distributed state, and coordinating GPU transfers across models and stages.

\sys{} targets these challenges as a cost-aware runtime that adapts RL post-training jobs into efficient execution plans. Its low-overhead controller selects a memory-feasible global plan and admits the transition using a cost model calibrated against the running job.
To estimate and execute a transition, \sys represents the distributed state of each replica of a model-stage (one model in one stage) as an Elastic Model Unit. It then employs a global transition graph to order the transformations and GPU transfers of these units. In a trace built from real data,
online TP/PP adaptation reduces average step latency by 27.7\% relative to the initial fixed TP/PP layout with DP scaling. In a 1{,}000-step run reaching 1{,}024 GPUs, six transitions consume 0.079\% of total run time. \sys improves end-to-end 8B PPO throughput by \mbox{2.14--7.27$\times$} over \mbox{OpenRLHF} and by \mbox{1.10--1.47$\times$} over Verl across
\mbox{diverse clusters}.
\end{abstract}


\settopmatter{printfolios=true,printacmref=false}
\maketitle
\thispagestyle{paperfirst}
\pagestyle{paperheader}
\enlargethispage{-95pt}

\section{Introduction}
\label{sec:introduction}

\begin{figure*}[t]
  \centering
  \begin{subfigure}[b]{0.48\textwidth}
    \centering
    \includegraphics[width=\linewidth]{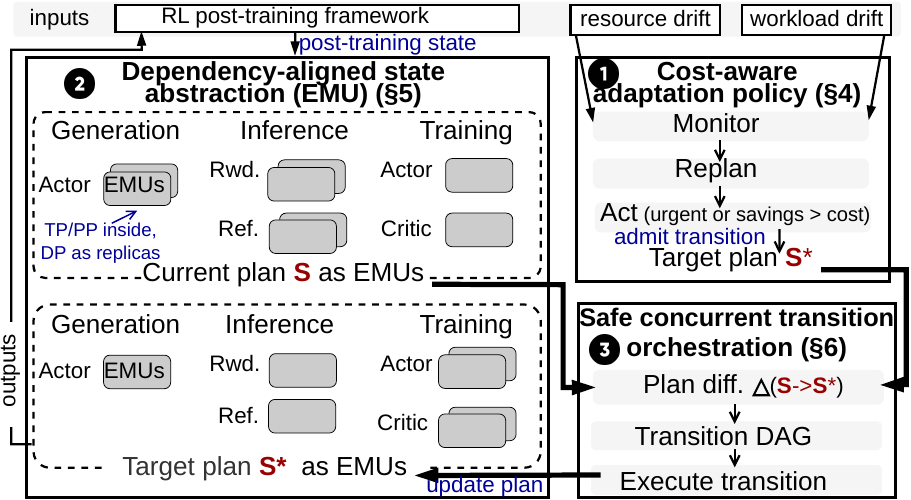}
    \caption{Components of \sys{}}
    \label{fig:archi-overview}
  \end{subfigure}
  \hfill
  \begin{subfigure}[b]{0.50\textwidth}
    \centering
    \includegraphics[width=\linewidth]{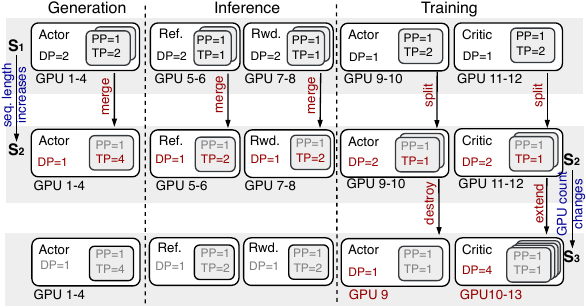}
    \caption{Two plan transitions on one job}
    \label{fig:archi-example}
  \end{subfigure}
  \caption{\sys{} overview.
  (a) \protect\myc{1} The controller replans on drift and
  admits the transition to $\mathcal{S}^*$ only if the current plan is infeasible or
  the savings repay the transition cost (\S\ref{sec:control}).
  \protect\myc{2} Both plans are represented as \abss{}, one per model-stage
  replica, with TP/PP inside the unit and DP as the replica count
  (\S\ref{sec:abstraction}).
  \protect\myc{3} The transition engine compiles the plan difference into a global transition DAG, adding resource-dependency edges when transient GPU overlap blocks an acquisition
  (\S\ref{sec:transition}).
  (b) Sequence-length growth reshards the five model-stages, with
  \code{Merge} in generation and inference, and \code{Split} in training
  ($\mathcal{S}_1\!\rightarrow\!\mathcal{S}_2$).
  A change in the GPU count removes an actor-training replica with \code{Destroy}
  and adds critic replicas with \code{Extend}
  ($\mathcal{S}_2\!\rightarrow\!\mathcal{S}_3$).}
  \label{fig:archi}
\end{figure*}

Reinforcement learning (RL) post-training is an important part of developing large language models (LLMs)~\cite{ouyang2022training,deepseekai2025deepseekr1}.
Each run is a distributed workload that coordinates an actor with the critic, reward, and reference models required by the RL algorithm across
generation, inference, and training
stages~\cite{hu2024openrlhf,sheng2024hybridflow,fu2025areal,lei2024puzzle}.
A \emph{model-stage} is one model assigned to one stage (\eg actor generation).
An \emph{execution plan} assigns GPUs to the model-stages and sets their data, tensor,
and pipeline parallelism (DP/TP/PP) degrees~\cite{mei2025real}; it also
determines state placement, communication, and the division of work.

One RL post-training run can use 100{,}000 GPU-hours~\cite{khatri2025art}.
During that run, GPUs may fail or be revoked~\cite{tenplex,wu2024can}, and cluster
schedulers may reassign nodes~\cite{li2023easyscale}.
Moreover, network contention can slow down
communication~\cite{saurabh2020congestion}.
At the same time, workload characteristics change as the actor learns:
our measurements show a 16$\times$ increase in generated sequence length within 1{,}000
steps (\S\ref{sec:need}), reflecting how actors learn to produce longer
trajectories~\cite{deepseekai2025deepseekr1}.
This growth increases memory pressure, which can render the current execution plan
infeasible and shift bottlenecks among generation, inference, and
training stages~\cite{zhong2025rlhfuse,sheng2024hybridflow}.
Consequently, static provisioning faces a trade-off:
allocating resources early for the final sequence length wastes resources, whereas provisioning only
for the initial length risks running out of memory later.
The set of changes in resource supply, workload demand, and achieved hardware efficiency is called \emph{drift}.

Existing systems only partially address this problem.
Several RL post-training frameworks support large spaces of execution plans, but keep their allocation and parallelism choices fixed after startup, even as different stages become bottlenecks~\cite{sheng2024hybridflow,hu2024openrlhf,fu2025areal,lei2024puzzle}.
Elastic training systems can change a running plan, but typically manage just one
model~\cite{li2023easyscale,mo2024heet,tenplex,jang2023oobleck}.
At a coarse granularity, checkpoint systems preserve state for failure recovery~\cite{wang2023gemini}, whereas
checkpoint-based resharding restarts the job and reconstructs state through host memory or
storage~\cite{pytorch_dist_ckpt,megatron_dist_ckpt,lian2025universal,wan2025bytecheckpoint}.
At a finer granularity, single-model state-management systems instead expose one model or its shards as the
state unit~\cite{tenplex,jang2023oobleck}.
In RL post-training, DynaRL selects resource assignments across stages
only within a fixed pool and executes them through per-component migrations~\cite{wang2026dynarl}.
Coordinating coupled RL models leaves three questions open: \textbf{when} to adapt, \textbf{what} state to reuse, and \textbf{how} to transition.

First, deciding \emph{when} to transition requires selecting a global plan and verifying that its expected savings outweigh the transition cost, accounting for all coupled model-stages.
Second, determining \emph{what} to reuse is difficult because current and target plans share logical state (parameters and optimizer state) but differ in sharding, placement, and replication.
While transitions can reuse GPU-resident state~\cite{tenplex,jang2023oobleck}, the choice of state unit involves a granularity trade-off: a coarse unit simplifies planning but transfers redundant state, whereas a fine unit minimizes transfer but leaves complex shard-level constraints to the planner.
Third, executing \emph{how} to transition is highly constrained: when no free GPUs remain, one model-stage must release resources before another can expand.
The runtime must carefully order these cross-stage dependencies while preserving each model-stage's logical state and respecting GPU memory limits.

These challenges jointly motivate \sys{}, an online, cost-aware runtime that dynamically adapts execution plans and distributed state for RL post-training on GPU clusters.
Our key insight is that plan adaptation becomes tractable when the state boundary matches the job's dependencies.
This state boundary allows \sys{}'s transition engine to execute plan changes directly, without reconstructing the full job state.
Specifically, this work establishes the following key contributions, one for each question.

\mypar{(1) Low-overhead, cost-aware adaptation policy}
We propose a control policy that decides \emph{when} to adapt under workload drift and resource volatility.
By combining lightweight, event-driven triggers with an online-calibrated cost model, the policy identifies memory-feasible target plans and admits transitions when expected performance gains outweigh estimated transition overhead (\S\ref{sec:control}).
The policy also complements fault-tolerance systems, dynamically replanning from intact units during sudden node failures.

\mypar{(2) A dependency-aligned state abstraction via Elastic Model Units}
We introduce the \emph{Elastic Model Unit} (\abs{}), a principled state abstraction that defines \emph{what} state to reuse by aligning state boundaries with model dependencies.
This abstraction encapsulates intra-model tensor and pipeline parallelism inside the unit, while exposing data-parallel replicas for flexible scaling.
We define four core state-transformation primitives that cover all parallel configurations in the plan space while preserving training semantics (\S\ref{sec:abstraction}).

\mypar{(3) Safe, concurrent transition orchestration}
We design an execution protocol that coordinates \emph{how} to transition across coupled models under tight cluster resources.
The transition engine compiles global plan changes into a directed acyclic graph (DAG) of \abs{} primitives to maximize safe, concurrent state transfers (\S\ref{sec:transition}).
To avoid deadlocks in memory-constrained environments, the protocol dynamically prioritizes resource-releasing operations without blocking independent concurrent transfers.

\sys{} is implemented in 43k lines of Python, C++, and CUDA/HIP, integrating with vLLM, DeepSpeed, and Megatron-LM, using standard NCCL/RCCL collectives.
Across the evaluated clusters, \sys{} improves end-to-end 8B PPO throughput by 2.14--7.27$\times$ (median 3.99$\times$) over OpenRLHF and by 1.10--1.47$\times$ over Verl.
Its selected plans stay within 5\% of the empirical optimum in all 18 measured settings.
In a trace built from real data (\S\ref{subsec:runtime-drift}), online TP/PP adaptation reduces average step latency by 27.7\% relative to a fixed layout with DP scaling.
Across three held-out traces, \sys{} averages 858.7\unit{s}/step compared to 928.3\unit{s} for DynaRL-style admission.
Scaling with \abss{} is 3.8--16.2$\times$ faster than with Oobleck and Tenplex, and coordinated transitions succeed in all overlap trials, compared to only 34--62\% for DynaRL's per-component migrations.
Finally, six transitions during a 1{,}000-step run scaling to 1{,}024 GPUs consume just 0.079\% of total execution time.

\section{Design Overview}
\label{sec:overview}

\sys{} separates its low-overhead adaptation policy from transition orchestration
(\F\ref{fig:archi-overview}).
The controller's monitor reads sequence length, available GPUs, peak memory, and achieved
compute and communication efficiency.
The planner selects a memory-feasible target $\mathcal{S}^*$ with the lowest
predicted steady-state step latency. Admission checks feasibility and whether the savings repay the one-time
transition cost estimated by the transition engine (\S\ref{sec:control}).

Because several models appear at multiple stages of an RL post-training step, reusing
GPU-resident state calls for per-model-stage state units and coordination across model-stages that share GPUs.
\sys{} therefore represents the current and target plans as collections of Elastic Model Units (\abss{}, \S\ref{sec:abstraction}).
Each \abs{} contains the tightly coupled TP/PP layout and state of one model-stage
replica, and the number of replicas sets the DP degree.
The engine converts the plan difference into \code{Split}, \code{Merge}, \code{Extend}, and \code{Destroy}
primitives, orders their dependencies in a global transition DAG (\S\ref{sec:transition}), and executes it after admission.
\F\ref{fig:archi-example} shows two such transitions, triggered by
sequence-length growth and GPU-count changes.

\section{Background and Design Space}

RL post-training couples model-stages whose resource needs and performance change
during a run. Adapting the execution plan therefore requires a global view.

\subsection{RL Post-Training Execution Model}
\label{sec:background}

\begin{figure}[t]
	\centering
	\includegraphics[width=0.38\textwidth]{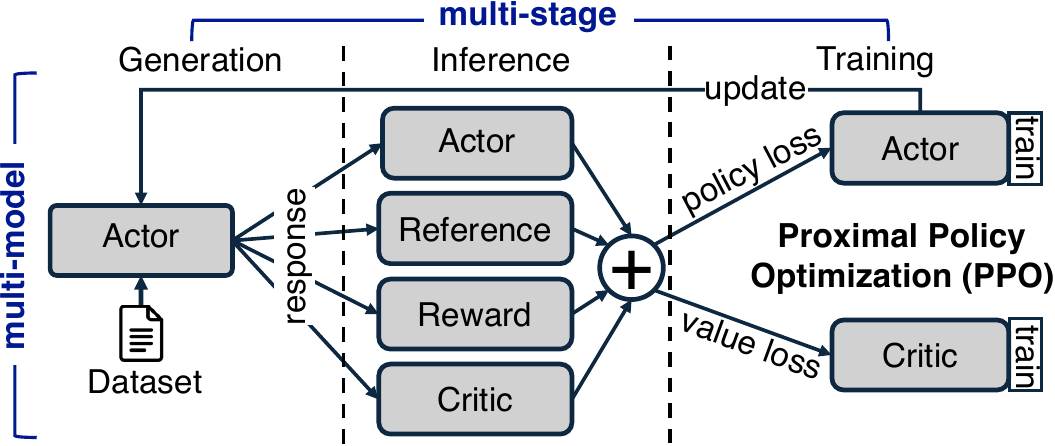}
	\caption{Multi-model, multi-stage workflow of RL post-training with Proximal Policy Optimization (PPO).}
	\label{fig:rlhf}
\end{figure}

RL post-training algorithms such as Proximal Policy Optimization (PPO)~\cite{schulman2017ppo},
ReMax~\cite{li2024remax}, and Group Relative Policy Optimization (GRPO)~\cite{shao2024deepseekmath} run several models across the three stages.
Take PPO (\F\ref{fig:rlhf}) as an example~\cite{ouyang2022training}.
Its actor generates responses, the reward model scores them, the critic estimates
their values, and a frozen reference model constrains policy updates.
Generation produces responses, and inference computes their probabilities, values, and
rewards. Training updates the actor and critic from these signals.
As a result, a model can have different compute and memory needs across the three stages.

\mypar{Execution plans and coupling}
The execution plan defined in \S\ref{sec:introduction} couples these model-stages
through their GPU assignments and parallelism.
DP runs replicas on different data, TP partitions tensors within a layer, and PP
partitions layers into sequential pipeline stages~\cite{shoeybi2020megatronlm,rajbhandari2020zero}.
For example, raising TP for actor generation changes how actor weights must be
resharded after training. Accelerating the actor has little effect if
the critic remains the bottleneck.
Such coupling exists in both synchronous and asynchronous frameworks.
Synchronous systems such as Verl (HybridFlow)~\cite{sheng2024hybridflow} leave workers
waiting at a barrier when stages are imbalanced. Asynchronous systems such as AReaL~\cite{fu2025areal}
instead show mismatched rollout and update rates.

Coupling also constrains transition execution.
Serialized model-stages may share GPUs, but concurrent model-stages in a valid plan use disjoint
GPU sets.
When the job has no free GPUs, one model-stage may need to release GPUs
before another can expand.
A transition must therefore respect
transient GPU ownership across model-stages, which the global transition DAG encodes.

\subsection{Sources of Drift}
\label{sec:need}

The best plan can change during a run: a static plan can become inefficient or infeasible when resource supply, workload
demand, or achieved hardware efficiency changes.

\begin{figure}[t]
	\centering
	\begin{minipage}{\linewidth}
		\centering
		\begin{subfigure}[b]{0.32\linewidth}
			\centering
			\includegraphics[width=\linewidth]{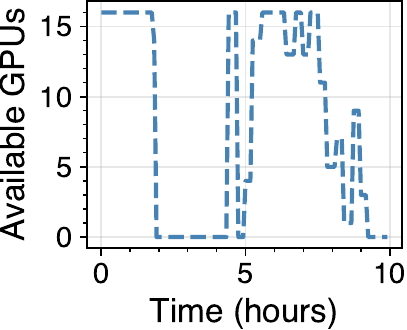}
			\caption{\code{us-east-2b} zone}
			\label{fig:gpu-ava1}
		\end{subfigure}
		\hfill
		\begin{subfigure}[b]{0.32\linewidth}
			\centering
			\includegraphics[width=\linewidth]{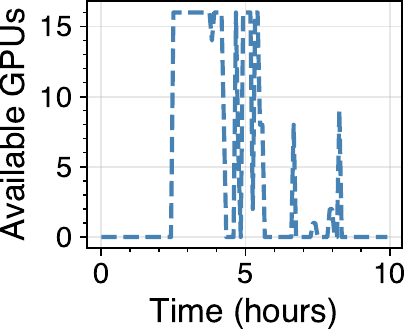}
			\caption{\code{us-west-2a} zone}
			\label{fig:gpu-ava2}
		\end{subfigure}
		\hfill
		\begin{subfigure}[b]{0.32\linewidth}
			\centering
			\includegraphics[width=\linewidth]{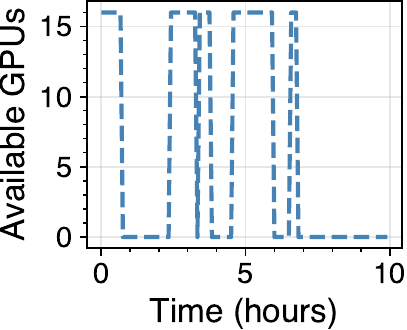}
			\caption{\code{us-west-2c} zone}
			\label{fig:gpu-ava3}
		\end{subfigure}
		\caption{Frequent GPU availability fluctuations in shared cloud environments (\code{p3.2xlarge} nodes) over ten hours.}
		\label{fig:gpu-ava}
	\end{minipage}
\end{figure}

\begin{figure}[t]
	\centering
	\begin{minipage}{1\linewidth}
		\centering
		\begin{subfigure}[b]{0.43\linewidth}
			\centering
			\includegraphics[width=1\textwidth]{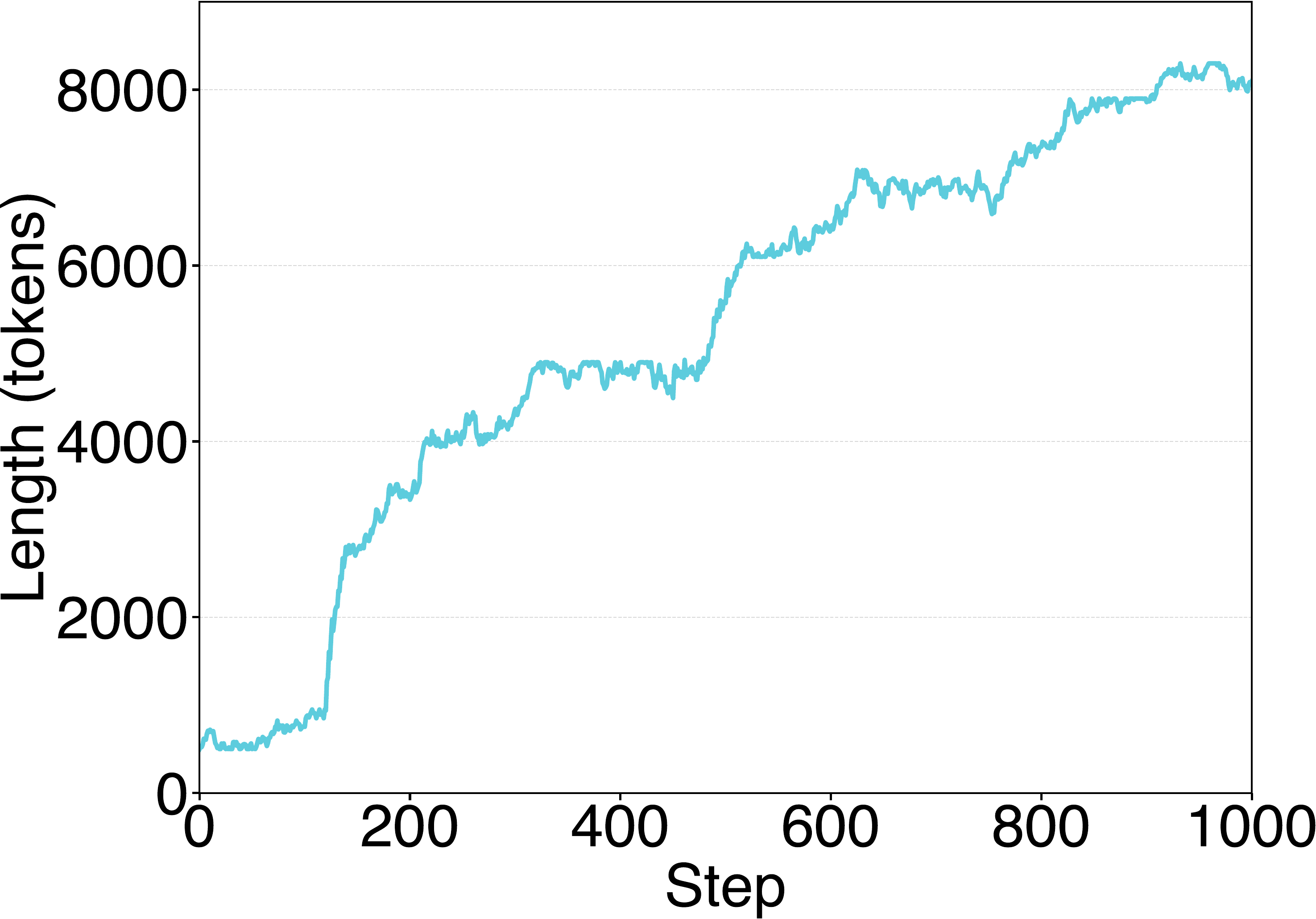}
			\caption{Sequence-length growth}
			\label{fig:sequence-length}
		\end{subfigure}
		\begin{subfigure}[b]{0.52\linewidth}
			\centering
			\includegraphics[width=1\linewidth]{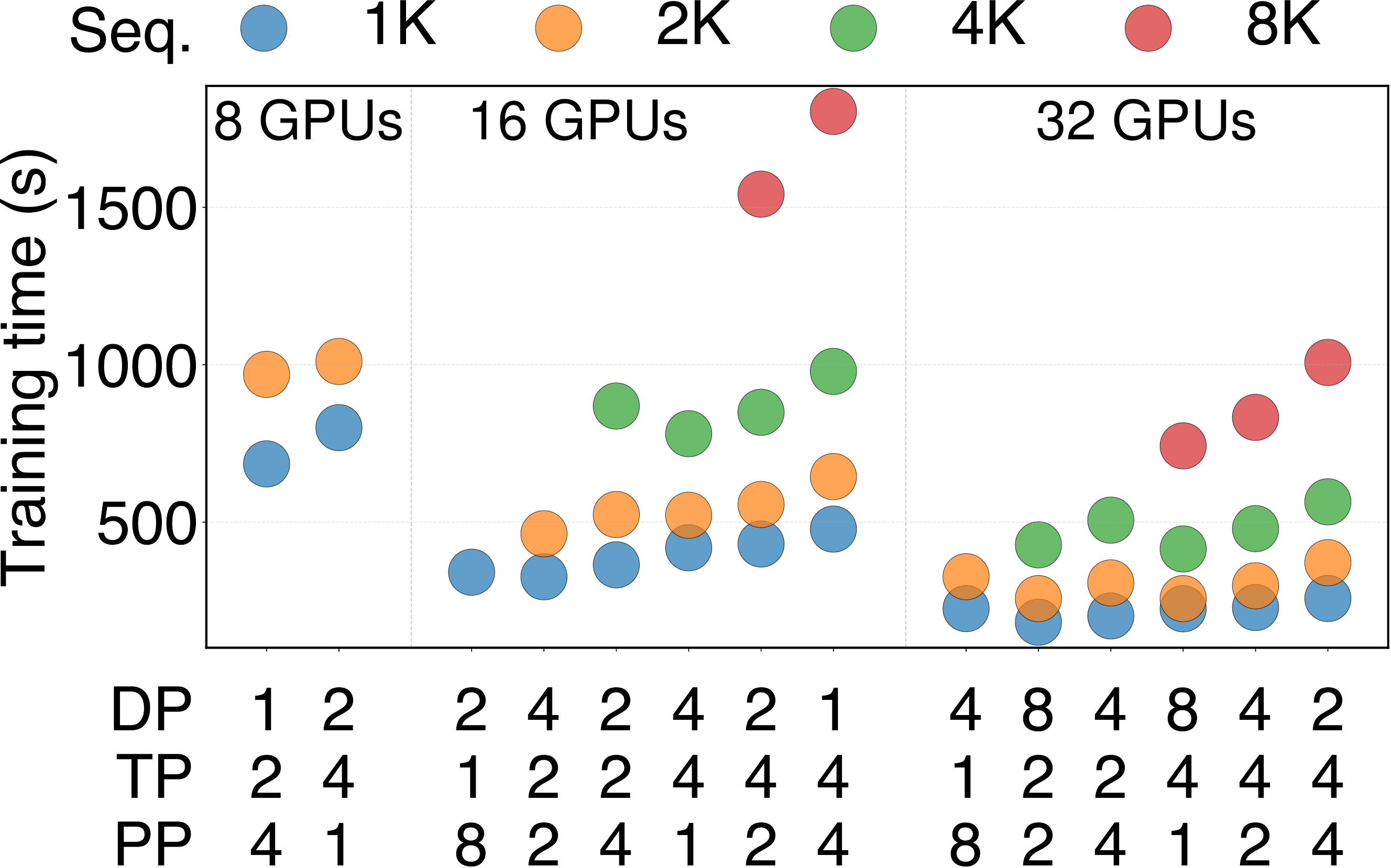}
			\caption{Best plan shifts with length}
			\label{fig:sequence-length-training-time-compare-8b}
		\end{subfigure}
		\caption{Generated sequence length grows during training (a), changing which measured training plan is fastest (b). Missing data points indicate out-of-memory (OOM) errors.}
		\label{fig:sequence-length-scaling}
	\end{minipage}
\end{figure}

\begin{table*}[!t]
	\centering
	\footnotesize
\caption{Approaches to online RL post-training adaptation along the three design decisions of \S\ref{sec:design-space}:
\emph{when} to adapt (timing), \emph{what} state to reuse (the boundary), and \emph{how} to transition. \sys{} uses a dependency-aligned model-stage boundary.}
	\label{tab:rlhf-reconfiguration-extended}
	\renewcommand{\arraystretch}{1.02}
	\setlength{\tabcolsep}{4pt}
	\begin{tabular}{@{}>{\centering\arraybackslash}p{0.3cm}
		>{\raggedright\arraybackslash}p{2.9cm}
		>{\raggedright\arraybackslash}p{4.95cm}
		>{\raggedright\arraybackslash}p{2.4cm}
		>{\raggedright\arraybackslash}p{3.1cm}
		>{\raggedright\arraybackslash}p{2.7cm}@{}}
		\toprule
		&
		\textbf{Approach} &
		\textbf{Representative systems} &
		\textbf{Timing (\emph{when})} &
		\textbf{Boundary (\emph{what})} &
		\textbf{Execution (\emph{how})} \\
		\midrule

		\protect\myc{A} &
		Fixed-plan RL post-training execution &
		TRL~\cite{vonwerra2022trl}, DeepSpeed-Chat~\cite{yao2023deepspeed};\newline
			Verl~\cite{sheng2024hybridflow}, AReaL~\cite{fu2025areal}, ReaL~\cite{mei2025real}, PUZZLE~\cite{lei2024puzzle},
			RLHFuse~\cite{zhong2025rlhfuse}, ROLL~\cite{wang2025roll}, OpenRLHF~\cite{hu2024openrlhf} &
		none: plan fixed at startup &
		-- &
		-- \\

		\midrule

		\protect\myc{B} &
		Checkpoint/restart &
		DCP~\cite{pytorch_dist_ckpt}, MCP~\cite{megatron_dist_ckpt}, UCP~\cite{lian2025universal}, GCR~\cite{zeng2026gcr};\newline
			Gemini~\cite{wang2023gemini}, ByteCheckpoint~\cite{wan2025bytecheckpoint} &
		on failure or restart &
		checkpoint state &
		checkpoint + restart \\

		\midrule

		\protect\myc{C} &
		Single-model state~management &
		Tenplex~\cite{tenplex}, Oobleck~\cite{jang2023oobleck} &
		on a resource change &
		selective single-model state &
		host/GPU state~transfer \\

		\midrule

		\protect\myc{D} &
		Dynamic scheduling in a fixed pool &
		DynaRL~\cite{wang2026dynarl} &
		utilization and predicted throughput &
		per-component state &
		global allocation with per-component migration \\

		\midrule\midrule

		&
		Dependency-aligned elastic multi-model state~management &
		\textbf{\sys} &
		payback-based admission &
		dependency-aligned model-stage replica (\abs{}) &
		DAG of four primitives over GPU-direct links where available \\

		\bottomrule
	\end{tabular}
\end{table*}

\mypar{Resource supply}
The total number of GPUs available to a job can change during a run.
Spot instances may be reclaimed, and shared-cluster schedulers may reallocate
nodes~\cite{wu2024can,duan2024parcae,li2023easyscale,mo2024heet,whitton2025hpc}.
An analysis of AWS traces~\cite{wu2024can} found more than ten availability changes
for a 16-GPU job over ten hours (\F\ref{fig:gpu-ava}).
In our plan enumeration for Llama-3.1-8B, the best TP/PP layout differs between 32 and 256
GPUs. A resource change can therefore call for a different plan, not just a resize.

\mypar{Workload demand}
As training progresses, actors can generate longer trajectories~\cite{deepseekai2025deepseekr1},
increasing the key-value (KV) cache footprint during generation and the activation
memory during training.
For Llama-3.1-8B~\cite{grattafiori2024llama3herdmodels}, generated sequence length
grows from 500 to 8{,}000 tokens within 1{,}000 steps
(\F\ref{fig:sequence-length}).
At 16 GPUs, the best (DP, TP, PP) training plan changes with this growth. \code{(4,2,2)} is
fastest at 2K tokens but runs out of memory at 4K. \code{(4,4,1)} remains feasible
at 4K but adds communication overhead at 2K
(\F\ref{fig:sequence-length-training-time-compare-8b}).

\mypar{Achieved hardware efficiency}
Network congestion, thermal throttling, and multi-tenant interference change the
step latency of a plan even when its allocation and workload remain fixed~\cite{saurabh2020congestion,darzi2025host,xavier2016interference}.
These effects can shift the best plan and make calibration against measured execution useful. Offline estimates
can also be wrong: for the Llama-3.1-8B model, an offline predictor~\cite{llm-analysis} selects an infeasible
plan at 8 GPUs.
At 64--256 GPUs, its selected plans are up to 1.56$\times$ slower than the best measured plans.
\sys{} therefore calibrates its compute, communication, and memory estimates against the
running job (\S\ref{subsec:monitor}--\S\ref{subsec:cost}).

Drift affects model-stages differently, so adaptation must cover the whole job
rather than resize one model-stage.
In \F\ref{fig:archi-example}, longer sequences change the five model-stages, with different resharding
directions for generation and training, while a change in GPU count
removes an actor-training replica and adds critic replicas.
Training can use \code{Split} to lower TP only while it has memory headroom, and \code{(4,2,2)} has no headroom at 4K
(\F\ref{fig:sequence-length-training-time-compare-8b}).

\subsection{Design Space for Online Adaptation}
\label{sec:design-space}

Online adaptation links three decisions.
First, \textbf{when} to adapt depends on whether a feasible target plan exists under the
current resource and memory constraints, and whether its global
step-latency savings can repay the transition cost before conditions change again~\cite{qiao2021pollux,jayaram2023sia}.
Second, the state boundary determines \textbf{what} state can be reused or must be
moved.
A coarse boundary moves job-wide state even for a localized change~\cite{lian2025universal,wan2025bytecheckpoint}, while a shard-level boundary
requires coordinating shard routing and collective synchronization~\cite{tenplex}.
Finally, \textbf{how} the transition executes determines its cost. State transfers should use
GPU-direct paths where possible, and primitives that contend for the same GPUs
must be ordered~\cite{tenplex,jang2023oobleck,mai2020kungfu,thorpe2023bamboo}.
Prior approaches make these decisions per job, model, or component (\T\ref{tab:rlhf-reconfiguration-extended}), and even replanning without payback-based admission trails \sys{} (\S\ref{subsec:runtime-drift}).
\sys{} addresses all three with a cost-aware adaptation policy (\S\ref{sec:control}), a dependency-aligned state abstraction (\S\ref{sec:abstraction}), and safe concurrent transition orchestration (\S\ref{sec:transition}).

\section{Cost-Aware Adaptation Policy}
\label{sec:control}

The controller decides when to adapt: it replans on resource or memory events or threshold crossings and admits an executable transition if the current plan is infeasible (urgent) or the savings repay the transition cost (opportunistic).

\subsection{Monitor Phase: Detecting Runtime Drift}
\label{subsec:monitor}

The monitor reads sequence length, the available GPU pool
$P_{\mathrm{gpu}}$, peak memory, and achieved compute and communication efficiencies
to track drift and calibrate the cost model (\S\ref{subsec:cost}).
Ray and NVML (or ROCm SMI)~\cite{moreritz2018ray,nvidia-nvml,amd-rocm-smi} provide the hardware inventory. Framework instrumentation provides
kernel and collective times.

Since these signals behave differently, the monitor uses two kinds of triggers.
A change in $P_{\mathrm{gpu}}$ or a predicted memory violation triggers replanning.
A drifting signal $x$ triggers replanning when it differs from its reference value (its value at the last replan) by more
than a relative threshold $\delta_x$.
All experiments use \mbox{$\delta_{\mathrm{len}}=30\%$} for per-step sequence length.
Reference values are reset after every replan, even when the candidate transition is rejected, so a persistent deviation does not retrigger at every step.
Either trigger starts a CPU-side replan when any running transition ends, or at once if GPUs are lost (\S\ref{subsec:controller}).
The admission rule in \S\ref{subsec:controller} then decides whether to
execute the transition.

\sys{} adapts at safe boundaries: RL-step completion in synchronous execution and
weight synchronization in asynchronous execution. The RL framework schedules weight synchronization and handles in-flight rollouts, which keep their policy versions. Affected \abss{} first complete
in-flight accesses, collectives, and optimizer updates. The transition then
carries parameter tensors, optimizer tensors, update counters, and RNG and dataloader
state into the target layout, preserving model-stage versions, policy versions, and the framework's sample-consumption rules.

\subsection{Replan Phase: Selecting a Target Plan}
\label{subsec:cost}

Let $\mathcal{M}$ be the set of model-stages in the RL job.
A plan assigns every model-stage $\mathsf{M}\in\mathcal{M}$ a candidate
$\pi_{\mathsf{M}}=\langle \mathrm{tp}_{\mathsf{M}},\allowbreak\mathrm{pp}_{\mathsf{M}},\allowbreak
\mathrm{dp}_{\mathsf{M}},\allowbreak\mathcal{R}_{\mathsf{M}}\rangle$, consisting of its TP, PP, and DP degrees and assigned GPU set $\mathcal{R}_{\mathsf{M}}$.
At the controller level, the global plan is $\mathcal{S}=\{\pi_{\mathsf{M}}\mid
\mathsf{M}\in\mathcal{M}\}$.
The planner seeks the lowest-latency plan that fits within the memory and GPU-pool constraints:
\begin{equation}
	\begin{aligned}
		\mathcal{S}^* = {}&\!\arg\min_{\mathcal{S}} \quad \widehat L_{\mathrm{step}}(\mathcal{S}) \\
		\text{s.t.} \quad & \mathrm{Mem}_{\mathrm{peak}}(\mathcal{S}) \le \mathrm{Mem}_{\mathrm{cap}}, \\
		& \sum\nolimits_{\mathsf{M} \in \mathcal{C}_s} |\mathcal{R}_{\mathsf{M}}| \le |P_{\mathrm{gpu}}|, \quad \forall s \in \{\code{Gen}, \code{Inf}, \code{Train}\}.
	\end{aligned}
	\label{prob:control}
\end{equation}
Here, $\mathcal{S}^*$ is the target plan, and
$\widehat L_{\mathrm{step}}(\mathcal{S})$ is the predicted end-to-end RL-step latency.
$\mathrm{Mem}_{\mathrm{peak}}(\mathcal{S})$ is the largest predicted per-GPU memory
footprint over the plan's execution schedule. The first constraint bounds it by the
per-GPU memory capacity $\mathrm{Mem}_{\mathrm{cap}}$.
The second constraint bounds the total GPU allocation of $\mathcal{C}_s$ by the pool size, where $\mathcal{C}_s$
contains the model-stages that run concurrently during stage $s$ on disjoint GPU
sets.
Solving Eq.~(\ref{prob:control}) online requires three components: a
cost model for the objective and constraints, online calibration, and a search fast enough to run at every replan.

\mypar{Cost model}
\sys{} extends the cost-model-guided planning of NanoFlow and Alpa~\cite{zhu2025nanoflow,zheng2022alpa}
to coupled multi-model plans.
For each model-stage candidate, the cost model estimates the local computation, pipeline bubbles, and
the tightly coupled TP/PP collectives of each replica.
It estimates computation time from peak compute, and collective time from the bandwidth
of each communication path.
The model-stage estimate adds DP synchronization and is determined by its slowest
replica.
At the job level, \sys{} derives an execution-overlap graph from the RL workflow,
and $\widehat L_{\mathrm{step}}$ is the critical-path latency of that graph.
The graph serializes model-stages that share GPUs and runs independent ones in parallel. These
parallel model-stages form the $\mathcal{C}_s$ of Eq.~(\ref{prob:control}).
For asynchronous execution, the critical path spans the interval between successive
weight synchronizations, which are the safe boundaries.

Peak memory covers parameters, gradients, optimizer state, activations, KV cache,
co-resident idle model-stages, and workspaces. Shapes, precision, sharding, batch size,
sequence length, and recomputation set these footprints~\cite{llm-analysis}.
Measured peaks (\S\ref{subsec:monitor}) calibrate them, and the feasibility check (\S\ref{subsec:controller}) uses the same estimates with 10\% headroom.

\mypar{Calibration}
\sys{} calibrates its cost model online without requiring a complete execution profile before startup. At startup,
operation shapes and sharding rules provide FLOP
counts, message volumes, and memory footprints. The hardware inventory provides peak
compute and link bandwidths. Because peak rates overstate what kernels and collectives achieve,
\sys{} learns the shortfall from measured execution times. It stores one efficiency
per operation class (\eg GEMM, attention, or an all-reduce on one link type) rather than per operation, so plans built
from the same classes share these efficiencies. Unseen classes start with conservative values and are refined after each step.
A new model can reuse estimates when its operation shapes and classes match.

\mypar{Plan search}
\sys{} first estimates each model-stage's DP/TP/PP candidates,
discarding those that exceed memory capacity or are dominated, to form a
latency--resource frontier. Candidates keep TP groups within one node when possible. Dynamic programming then selects one candidate per
model-stage under the plan-level memory bound and stage-wise GPU budgets to produce
$\mathcal{S}^*$, minimizing predicted steady-state latency.

Replanning runs in one CPU process on the head node, using no GPUs
or training nodes. \S\ref{subsec:cost-model-efficacy} reports calibration
fit, decision overhead, and replanning memory, and \S\ref{subsec:empirical-optima} evaluates selection quality by executing all 480
valid plans.

\subsection{Act Phase: Admitting a Transition}
\label{subsec:controller}

A replan yields a target plan $\mathcal{S}^*$.
\sys{} applies two criteria, \textit{feasibility} and \textit{profitability}, and executes an admitted transition only at a safe boundary (\S\ref{subsec:monitor}).

\mypar{Feasibility}
A transition is executable when, for its target plan and DAG, GPU memory can hold the retained state, workspaces,
intermediate units (including replicas created by \code{Split}, \S\ref{subsec:emu_primitives}), co-resident model state, and collective buffers. The controller performs this memory and GPU-pool
check before state movement. If $\mathcal{S}^*$ is rejected, \sys{} tries slightly slower candidates with cheaper transitions as the new $\mathcal{S}^*$ before keeping the current plan. Resource revocation, placement
change, or memory pressure that invalidates the current plan triggers the urgent
path, which runs the same check and preserves logical state
while waiving profitability. This path also considers smaller per-replica batches and recomputation, and falls back to checkpoint/restart if no candidate passes. \sys{} detects lost GPUs, aborts communicators that include the dead ranks, and replans from the intact units. Fault-tolerance systems~\cite{wang2023gemini,jang2023oobleck} restore state whose last copy is lost.
\par
\mypar{Profitability}
The transition engine estimates the critical-path cost $\widehat L_{\mathrm{tran}}$
of state movement and setup between the current and target \abs{} collections
(\S\ref{sec:abstraction}--\S\ref{sec:transition}).
Let $\Delta L=\widehat L_{\mathrm{step}}(\mathcal{S})
-\widehat L_{\mathrm{step}}(\mathcal{S}^*)$ be the predicted per-step
latency savings from switching the current plan $\mathcal{S}$
to the target plan $\mathcal{S}^*$. An executable opportunistic transition is admitted if
\begin{equation}
\Delta L>0
\quad\land\quad
\frac{\widehat L_{\mathrm{tran}}}{\Delta L}
\leq \gamma H_x.
\label{eq:profitability-gate}
\end{equation}
The ratio is the payback period: the steps needed for savings $\Delta L$ to repay
$\widehat L_{\mathrm{tran}}$. For the triggering signal $x$, $H_x$ estimates the steps
until its next threshold crossing (for $P_{\mathrm{gpu}}$, its next change). If several signals trigger at once, \sys{} uses the smallest $H_x$.
The confidence margin $\gamma\in[0,1]$ (0.5 by default) requires payback within a fraction of $H_x$ steps, reserving time for uncertainty. Setting $\gamma=0$ admits only urgent transitions.
\S\ref{subsec:runtime-drift} evaluates sensitivity to $\gamma$.

Each $H_x$ is an exponential moving average of past intervals between crossings, updated at
crossings rather than telemetry samples or admitted transitions.
Resetting the reference value
after every replan lets steady sequence-length growth produce repeated crossings
(\S\ref{subsec:monitor}). Before the first interval is observed, $H_x$ is one step.

The admission rule applies the payback principle of Pollux~\cite{qiao2021pollux} and
Sia~\cite{jayaram2023sia} to one coupled job.
Because resizing one model-stage can force changes to others, $\Delta L$ covers the
full RL step and $\widehat L_{\mathrm{tran}}$ covers the complete transition.
For example, a generation replica may be cheap to add unless its GPUs must come
from training. Releasing and resharding that training state can make the same
resize fail Eq.~(\ref{eq:profitability-gate}).

\section[Dependency-Aligned State Abstraction (EMU)]{Dependency-Aligned State\\Abstraction (\abs{})}
\label{sec:abstraction}

\begin{figure*}[t]
	\centering
	\includegraphics[width=\textwidth]{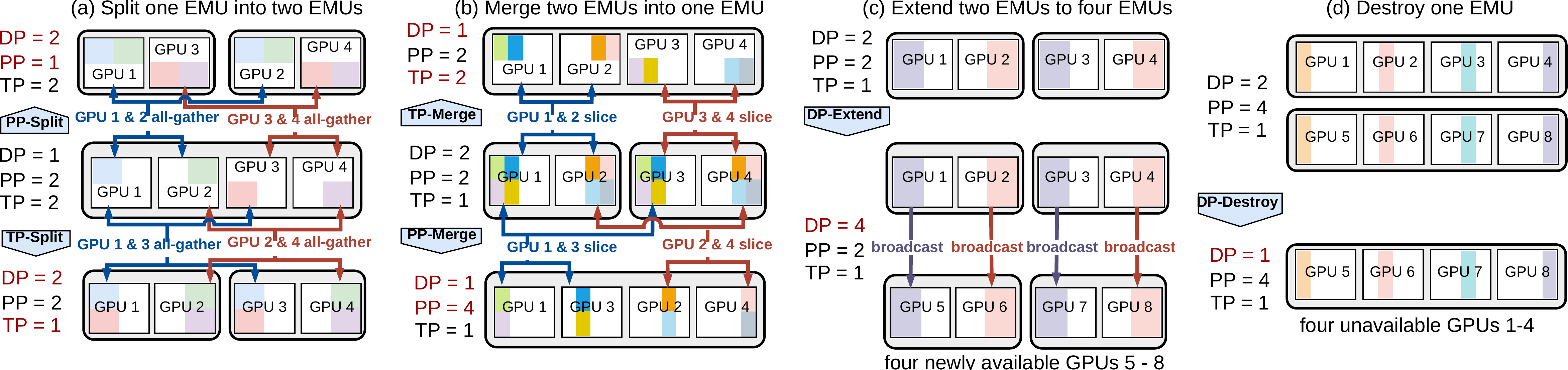}
	\caption{Core \abs{} primitives for online adaptation: \code{Split} (a) and \code{Merge} (b) reshard state by converting between tightly coupled TP/PP structure and loosely coupled replicas, while \code{Extend} (c) and \code{Destroy} (d) scale resources by creating or removing replicas. For the example in \S\ref{subsec:emu_def}, \code{Extend} adds replicas for critic training, and \code{Extend} followed by \code{Merge} increases TP for actor generation without changing its DP degree.}
	\label{fig:group}
\end{figure*}

To estimate and execute a candidate transition, the runtime needs a state boundary
that exposes what state to reuse without exposing every shard-level decision.

\mypar{Definition of \abs{}}
An \abs{} is one model-stage replica and the dependency-aligned state unit for online adaptation:
\textbf{\code{$\texttt{EMU}=\langle \mathsf{M}, \mathsf{T}_{\mathrm{tp} \times \mathrm{pp}}, \mathsf{R}, \boldsymbol{\theta}, \boldsymbol{\omega} \rangle$}}. Here, \code{$\mathsf{M}=(\mathit{model}, \mathit{stage})$} identifies the model-stage, \code{$\mathsf{T}_{\mathrm{tp} \times \mathrm{pp}}$} gives its TP/PP layout, and \code{$\mathsf{R}$} is its GPU set (\code{$|\mathsf{R}|=\mathrm{tp} \times \mathrm{pp}$}). Actor generation and actor training are thus separate model-stages, whose weights the RL framework synchronizes at safe boundaries. The parameter shards are \code{$\boldsymbol{\theta}=\{\theta_i\}_{i=1}^{\mathrm{tp} \times \mathrm{pp}}$}. For training, optimizer shards \code{$\boldsymbol{\omega}$} follow the same layout.
Neither \code{$\boldsymbol{\theta}$} nor \code{$\boldsymbol{\omega}$} is sharded across DP replicas (ZeRO-0).
DP is excluded from the internal structure of a single \abs{} (\S\ref{subsec:emu_def}), so the DP degree of model-stage \code{$\mathsf{M}$} is the number of its active \abss{}.
DP scaling thus only adds or removes units.
The planner reasons about whole units, while each primitive handles the underlying shards. \S\ref{subsec:implementation} lists the full logical state of a unit.

\mypar{Valid \abs{} collections}
An \abs{} collection is valid when each unit contains the shards required by its
TP/PP layout, concurrently executing units have disjoint GPU sets, and units with
the same model-stage index \code{$\mathsf{M}$} are DP replicas of the same logical state.

\mypar{Representing a global plan}
\S\ref{sec:control} expresses a plan as
$\mathcal{S}=\{\pi_{\mathsf{M}}\mid\mathsf{M}\in\mathcal{M}\}$, where each
$\pi_{\mathsf{M}}$ specifies TP, PP, DP, and a GPU allocation for one model-stage.
The runtime maps $\pi_{\mathsf{M}}$ to the collection
$E_{\mathsf{M}}=\Phi(\pi_{\mathsf{M}})=\{\texttt{EMU}_1,\texttt{EMU}_2,\dots,
\texttt{EMU}_{\mathrm{dp}_{\mathsf{M}}}\}$.
All units in the collection have the TP/PP layout specified by
$\pi_{\mathsf{M}}$, and their disjoint GPU sets together form the allocation
$\mathcal{R}_{\mathsf{M}}$.
The union
$\mathcal{E}(\mathcal{S})=\bigcup_{\mathsf{M}\in\mathcal{M}}
E_{\mathsf{M}}$ is the runtime representation of $\mathcal{S}$, and a transition turns it into $\mathcal{E}(\mathcal{S}^*)$ (\S\ref{sec:transition}).

\subsection{Why the Model-Stage Boundary}
\label{subsec:emu_def}

\begin{table}[t]
	\centering
	\footnotesize
	\setlength{\tabcolsep}{3.5pt}
	\caption{State boundaries for online adaptation, with the measured resource-scaling cost for an 8B actor/critic workload from 16 to 32 GPUs on Cluster~\#3 (the 16$\to$32 column of \T\ref{tab:reconf_overhead_cluster3}). The checkpoint row is measured with UCP~\cite{lian2025universal} and the shard-level row with Tenplex~\cite{tenplex}.}
	\label{tab:boundary-spectrum}
	\begin{tabular}{lccc}
		\toprule
		\textbf{Boundary} & \textbf{State managed} & \textbf{Planning unit} & \textbf{Cost} \\
		\midrule
		Checkpoint/restart & whole job & checkpoint & 836.74\unit{s} \\
		Shard level & shards & shard & 66.43\unit{s} \\
		Model-stage (\abs{}) & units & \abs{} & \textbf{6.52\unit{s}} \\
		\bottomrule
	\end{tabular}
\end{table}

A useful state boundary for RL post-training should satisfy four requirements. It should (i) confine state movement to the affected model-stage, (ii) let the
planner operate on replicas, (iii) preserve parameter and optimizer state at safe
boundaries (\S\ref{subsec:implementation}), and (iv) express DP/TP/PP changes with a small set of primitives
(\S\ref{subsec:emu_primitives}).

\mypar{The dependency structure determines the boundary}
In RL post-training, not all dimensions of distributed state and parallelism are coupled equally.
TP and PP require tightly synchronized execution.
Ranks exchange activations and gradients through collectives and point-to-point sends, so changing one shard's layout generally requires coordinated changes in the others.
DP is loosely coupled by comparison.
Replicas run the same computation and synchronize periodically, but one replica can be created, removed, or reassigned without restructuring the internal communication pattern of the others.
A job-level boundary can move unrelated state under localized drift, while a
shard-level boundary exposes tightly coupled TP/PP resharding decisions to the planner
(\T\ref{tab:boundary-spectrum}).
\sys{} keeps TP, PP, and their distributed state inside an \abs{} while leaving DP
outside, preserving each replica's internal synchronization and allowing adaptation at the
replica level.

The model-stage boundary also localizes state changes to the parts of the workflow
affected by the heterogeneous drift in \S\ref{sec:need}.
\abss{} are the state units over which DP/TP/PP choices are expressed, transformed, and evaluated online, while the controller (\S\ref{sec:control}) selects among these choices.

\mypar{Example}
Consider actor generation and critic training for 8B models initially sharing a
16-GPU pool, each configured as \code{(TP=1, PP=4, DP=2)},
\ie 8 GPUs per model-stage.
Suppose the sequence length increases from 2K to 4K and the pool grows to 32 GPUs.
The two models require different transitions. The actor increases TP from 1 to 2, to
provide memory for the larger KV cache during generation and avoid OOM errors. The critic increases DP from 2 to 4, to use the additional GPUs for training.
This reflects stage-specific bottlenecks: longer sequences increase memory pressure during actor generation, while critic training benefits more from replica-level throughput.
After the transition, the actor uses \code{(TP=2, PP=4, DP=2)}, and the critic uses \code{(TP=1, PP=4, DP=4)}.

The example combines an actor TP change with critic DP scaling. \T\ref{tab:boundary-spectrum}
isolates the resource-scaling cost for the same 16$\to$32-GPU workload
and compares this cost with checkpoint- and shard-level alternatives. Each measured cost reflects
a state boundary together with its primitives and transport implementation.

\subsection{Core \abs{} Primitives for Online Adaptation}
\label{subsec:emu_primitives}

Given the \abs{} boundary, \sys{} reduces online adaptation to four core primitives over \abss{}.
They cover two kinds of adaptation: \emph{parallelism resharding}, via \code{Split} and \code{Merge}, and \emph{resource scaling}, via \code{Extend} and \code{Destroy}.
Each primitive hides the underlying state movement (\eg collective communication) and exposes only its effect on the \abs{} collection.

(\textit{a}) \code{$\texttt{Split}(\texttt{EMU},\mathsf{T}_{\mathrm{sub}})\rightarrow\{\texttt{EMU}_1,\dots,\texttt{EMU}_k\}$}:
partitions one \abs{} with layout \code{$\mathsf{T}$} into \code{$k$} independent units with a smaller layout \code{$\mathsf{T}_{\mathrm{sub}}$}, \eg reducing \code{PP=2} to \code{PP=1} or \code{TP=2} to \code{TP=1} (\F\ref{fig:group}a).
It converts one tightly coupled unit into loosely coupled replicas by resharding
\code{$(\boldsymbol{\theta},\boldsymbol{\omega})$} across the same GPU set \code{$\mathsf{R}$}.

(\textit{b}) \code{$\texttt{Merge}(\{\texttt{EMU}_1,\dots,\texttt{EMU}_k\},\mathsf{T})\rightarrow\texttt{EMU}'$}:
fuses \code{$k$} replicas of the same model-stage state into a larger \abs{} with layout \code{$\mathsf{T}$}, \eg increasing \code{TP=1} to \code{TP=2} or \code{PP=2} to \code{PP=4} (\F\ref{fig:group}b).
It reshards the replicated state into one tightly coupled TP/PP unit.

(\textit{c}) \code{$\texttt{Extend}(\texttt{EMU},\mathsf{R}')\rightarrow\{\texttt{EMU},\texttt{EMU}'\}$}:
replicates an existing unit onto a newly allocated and disjoint GPU set \code{$\mathsf{R}'$}.
The new unit \code{\texttt{EMU}$'$} preserves the original model-stage index \code{$\mathsf{M}$} and layout \code{$\mathsf{T}$}, and holds replicated copies of \code{$\boldsymbol{\theta}$} and \code{$\boldsymbol{\omega}$}.
Each application increases DP by one, and extending both replicas raises \code{DP=2}
to \code{DP=4} (\F\ref{fig:group}c).

(\textit{d}) \code{$\texttt{Destroy}(\texttt{EMU})\rightarrow\emptyset$}:
terminates a redundant unit, releases its GPUs \code{$\mathsf{R}$}, and discards its
copy of \code{$(\boldsymbol{\theta},\boldsymbol{\omega})$} while another replica of the same model-stage retains the logical
state. This reduces DP by one, \eg from \code{DP=2} to \code{DP=1}
(\F\ref{fig:group}d).

\code{Split} and \code{Merge} are defined for any integer factor \code{$k$} between
compatible layouts. Repeated \code{Extend} and \code{Destroy} applications adjust
the DP degree by multiple units.
The planner searches power-of-two TP/PP degrees, following common practice on GPU clusters.

\mypar{The plan space}
The \abs{} representation uses homogeneous TP/PP \emph{within} a model-stage,
since all units sharing \code{$\mathsf{M}$} are replicas of one layout.
Different model-stages can use different layouts to match their computation and
memory requirements.
\S\ref{subsec:empirical-optima} shows that the selected plans stay within 5\% of the
empirical optimum in this plan space.

\mypar{Composing primitives}
Compositions of \code{Split}, \code{Destroy}, \code{Extend}, and \code{Merge} cover every transition in this plan space but require
intermediate memory and a schedule that retains each model-stage's logical state (\S\ref{sec:transition}).

\section{Safe Concurrent Transition Orchestration}
\label{sec:transition}

\begin{figure}[t]
	\centering
	\includegraphics[width=\columnwidth]{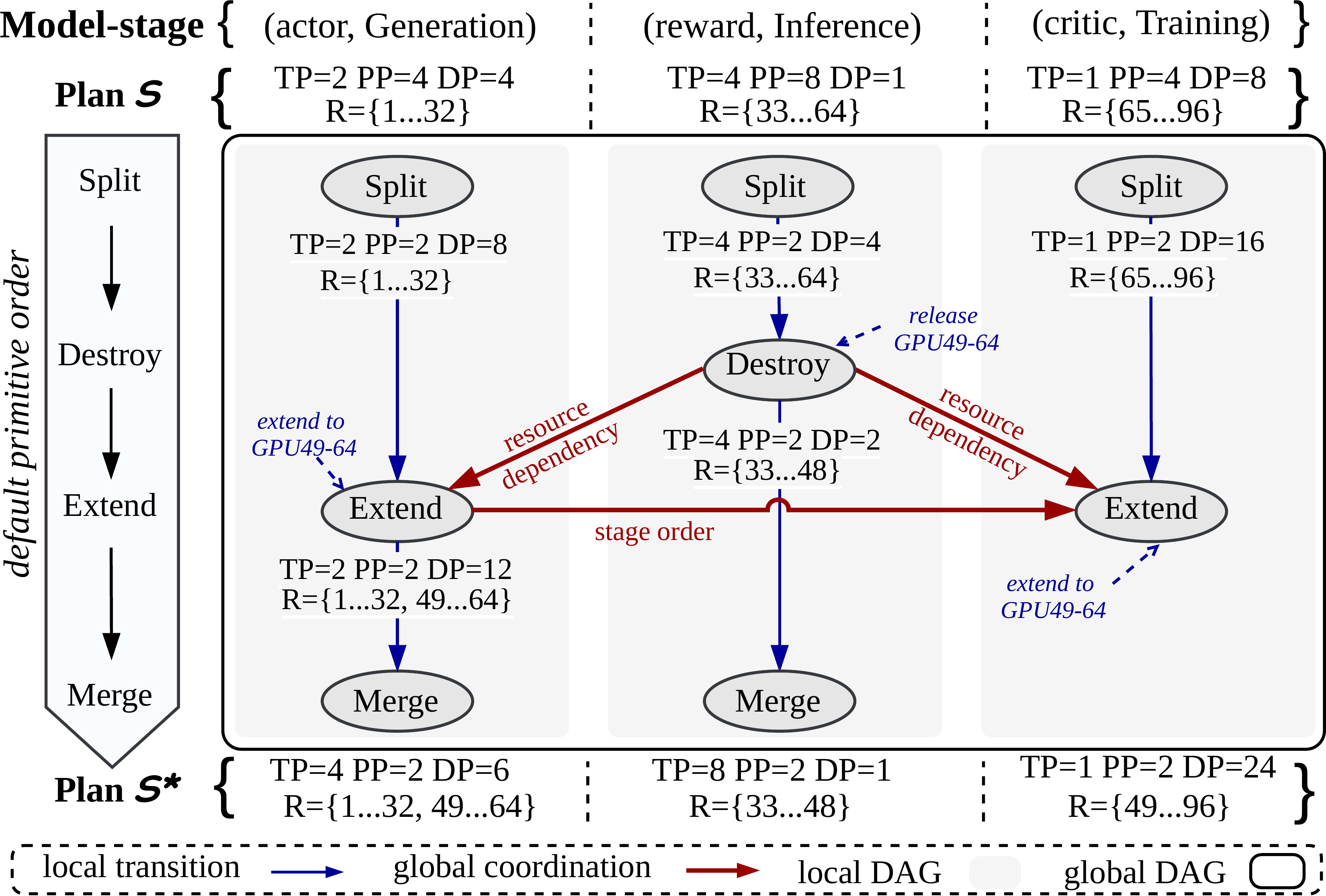}
	\caption{Constructing the global transition DAG from $\mathcal{S}$ to $\mathcal{S}^*$: \sys{} first translates each model-stage's change into a local DAG, then adds cross-model-stage dependency edges under transient GPU overlap.}
	\label{fig:dag}
\end{figure}

Starting from the \abs{} collections in \S\ref{sec:abstraction}, \sys{} plans how to transition per model-stage and adds cross-model-stage edges only when transient GPU overlap
blocks an acquisition. The resulting global transition DAG (\F\ref{fig:dag}) has a critical-path cost
$\widehat L_{\mathrm{tran}}$, used for admission (\S\ref{subsec:controller}) and estimated from bytes, link bandwidth, and communicator setup (\S\ref{subsec:cost}).

\subsection{Decomposition and DAG Construction}
\label{subsec:dag-construction}

The current and target global plans are
$\mathcal{S}=\{\pi_{\mathsf{M}}\mid\mathsf{M}\in\mathcal{M}\}$ and
$\mathcal{S}^*=\{\pi^*_{\mathsf{M}}\mid\mathsf{M}\in\mathcal{M}\}$.
Their runtime representations $\mathcal{E}(\mathcal{S})$ and $\mathcal{E}(\mathcal{S}^*)$ contain the corresponding \abs{} collections
$E_{\mathsf{M}}=\Phi(\pi_{\mathsf{M}})$ and
$E^*_{\mathsf{M}}=\Phi(\pi^*_{\mathsf{M}})$ defined in \S\ref{sec:abstraction}.
\sys{} summarizes the change for model-stage $\mathsf{M}$ as
$\Delta_{\mathsf{M}}=(E_{\mathsf{M}}\!\rightarrow\!E^*_{\mathsf{M}})$, which captures
TP/PP resharding, DP scaling, and GPU reassignment.
It then collects $\Delta=\{\Delta_{\mathsf{M}}\mid \mathsf{M}\in\mathcal{M}\}$, one subproblem per model-stage.

For each model-stage $\mathsf{M}$, \sys{} translates $\Delta_{\mathsf{M}}$ into a local DAG in the default order
\code{Split}{}$\,\rightarrow\,$\code{Destroy}{}$\,\rightarrow\,$\code{Extend}{}$\,\rightarrow\,$\code{Merge}.
It first computes the largest TP/PP
layout $\mathsf{T}_{\mathrm{sub}}$ into which both current and target units can be split.
The engine applies \code{Split} to obtain units with layout
$\mathsf{T}_{\mathrm{sub}}$. Next, \code{Destroy} removes redundant units and releases
their GPUs, and \code{Extend} adds replicas on the target GPUs. A unit with no other replica is destroyed only after \code{Extend} copies it. Finally,
\code{Merge} combines the intermediate units to obtain the target TP/PP
layout $\mathsf{T}^*$ and DP degree.
In \F\ref{fig:dag}, for reward inference, the engine splits the current \code{(TP=4, PP=8)} unit into four \code{(TP=4, PP=2)} units, destroys two, and merges the remaining two into the target \code{(TP=8, PP=2)}.

Valid local DAGs do not determine a global execution order when model-stages share
a near-capacity GPU pool. \sys{} combines local DAGs with two deterministic
rules: \textit{release-before-acquire} prioritizes resource-releasing primitives,
and \textit{stage order} breaks ties in generation~$\rightarrow$ inference~$\rightarrow$ training order.
In \F\ref{fig:dag}, actor and critic \code{Extend} wait for reward \code{Destroy} to release GPUs 49--64.
In \F\ref{fig:archi-example}, critic \code{Extend} onto GPU 10 waits for actor \code{Destroy}.
When an acquisition is blocked, the engine selects an executable
\code{Destroy} that releases the needed GPUs and inserts the corresponding
resource-dependency edge. \code{Destroy} primitives that depend, directly or transitively, on the blocked acquisition are not selected, so the DAG stays acyclic and free of circular waits (\S\ref{subsec:coordination-ablation}). If no such \code{Destroy} exists, the target plan fails the feasibility check. A blocked \code{Extend} waits for its target GPUs, while \code{Split}
retains its GPUs but runs only when its source and replica buffers pass the feasibility check.

\mypar{Control-plane efficiency}
Ordering primitives under state-transfer dependencies and resource requirements is a classic resource-constrained project scheduling problem~\cite{herroelen1998resource}.
\sys{} builds local DAGs in a fixed order, then greedily adds cross-model-stage edges.
\S\ref{subsec:transition-overhead} compares its planning time and schedule quality
with an off-the-shelf SCIP solver~\cite{hojny2025scip} on this transition-DAG scheduling problem.

\subsection{DAG Execution and Safety}
\label{subsec:implementation}

{\sloppy The global transition DAG orders \abs{} primitives according to their state and
resource dependencies. A primitive is ready only after its predecessors complete
and its resource requirements are satisfied. Ready primitives run concurrently
across GPU groups and streams.
\code{Split} uses position-wise \code{AllGather} within the smallest enclosing TP/PP group. \code{Merge} assembles the target layout from replicated state
(shards already resident on target ranks are not copied). \code{Extend} uses rank-aligned
\code{Broadcast}, and \code{Destroy} performs local teardown and memory release without collectives.\par}

\mypar{Safe concurrent execution}
Retained state, intermediate units, and collective
buffers must fit together in GPU memory at each operation: \code{Split} keeps source buffers live
until the collectives reading them complete, and \code{Destroy} releases memory
only after teardown. The
readiness, lifetime, and ordering rules coordinate concurrent state movement, while the
feasibility check enforces the per-GPU capacity bound (\S\ref{subsec:controller}). Since \code{Destroy} removes only redundant copies, a transition that stops partway loses no logical state unless the last copy is lost, and \sys{} replans from the intact units.

\mypar{Transport: sharded state over RDMA fabrics}
{\sloppy Inter-GPU state movement uses standard NCCL/RCCL collectives over communicators created for
the transition. \sys{} uses GPU-direct RDMA where available and introduces no custom
RDMA stack.
Intra-node movement usually uses NVLink or Infinity Fabric,
while inter-node movement uses RDMA-capable fabrics (InfiniBand, Slingshot) directly
between GPU buffers, without host staging.
Efficiency comes from how \sys{} matches these collectives to the sharding layout.
For \code{Extend}, \sys{} builds one communicator per rank pair at the same TP/PP position in the source unit and its new replica.
Each shard is sent once as a contiguous buffer to the corresponding destination
rank. Rank pairs transfer concurrently, subject to shared link bandwidth.\par}

\mypar{Resharding--transport interplay}
A transition is faster when target shard boundaries align with source ranks. When they do
not, state must be resharded across the network.
For compatible layouts, each resharding uses the largest common layout
$\mathsf{T}_{\mathrm{sub}}$. Layout changes then run locally after the required collectives.
Transfers remain contiguous.

\mypar{Preserving training semantics}
At a safe boundary, \code{Split} and \code{Merge} change placement and sharding
without changing logical state. An \abs{}'s logical state
$\Lambda=(\boldsymbol{\theta},\boldsymbol{\omega},c,v,p)$ includes parameters,
optimizer tensors, update counters $c$, model-stage and policy versions $v$, and the
minibatch position $p$ of the last completed update. \code{Extend} copies this state to a new DP
replica, and \code{Destroy} removes only a redundant copy. Besides $\Lambda$, \abss{} hold RNG and
dataloader state, which new replicas derive from their DP rank and $p$ in a global sample order, so no training sample is repeated or skipped. At synchronous step completion, no rollout is in flight, and KV caches and
gradient buffers hold no state. DP scaling preserves the
global batch size $\mathrm{dp} \times b \times a$ by adjusting $b$ and $a$, the per-replica
batch size and gradient-accumulation steps. The planner considers only DP degrees for which such $b$ and $a$ exist.
Fixed logical minibatches, loss normalization, sampling policy, and optimizer-step counts
preserve the update rule up to floating-point reduction order.

\section{Evaluation}
\label{sec:exp}

\begin{table*}[t]
\centering
\footnotesize
\setlength{\tabcolsep}{3pt}
\caption{Testbed GPU clusters. GCD denotes a graphics compute die, counted as one GPU. Bandwidths are nominal, and inter-node bandwidth is listed per link as send+receive.}
\label{tab:testbeds}
\begin{tabular}{lccc}
\toprule
\textbf{Property} & \textbf{Cluster \#1} & \textbf{Cluster \#2} & \textbf{Cluster \#3} \\
\midrule
\#Nodes / \#GPUs & 128 / 1{,}024 & 64 / 256 & 8 / 64 \\
GPUs per node
& 8$\times$ AMD MI250X GCDs (64\unit{GB})
& 4$\times$ NVIDIA A100 64\unit{GB}
& 8$\times$ NVIDIA H200 141\unit{GB} \\
Intra-node network
& Infinity Fabric (400 GB/s) & NVLink 3.0 (600 GB/s) & NVLink 4.0 (900 GB/s) \\
CPUs per node
& 1$\times$ 64-core AMD EPYC 7A53
& 1$\times$ 32-core Intel Xeon Platinum 8358
& 2$\times$ 32-core Intel Xeon Platinum 8562Y+ \\
Host memory & 512\unit{GB} & 512\unit{GB} & 2\unit{TB} \\
Inter-node network
& 4$\times$ HPE Cray Slingshot-11 (25+25 GB/s)
& 4$\times$ HDR100 InfiniBand (12.5+12.5 GB/s)
& 1$\times$ HDR InfiniBand (25+25 GB/s) \\
\bottomrule
\end{tabular}
\end{table*}

Q1 evaluates the complete system, Q2 contribution (1), Q3 contributions (2) and (3), and Q4 generality and the observed training behavior.

\noindent
\textbf{Q1.} Does \sys{} improve end-to-end RL post-training performance across model sizes and cluster scales? (\S\ref{sec:overall-performance})

\noindent
\textbf{Q2.} How well does the cost model fit measured training-stage latency and select target plans, and how does transition admission affect performance under drift? (\S\ref{sec:cost-model})

\noindent
\textbf{Q3.} Are \sys{}'s \abs{} primitives and transition engine efficient at cluster scale, and does cross-model-stage coordination avoid GPU conflicts? (\S\ref{sec:state-change})

\noindent
\textbf{Q4.} Do the performance benefits extend to other RL algorithms and execution modes, and does adaptation preserve the observed training behavior? (\S\ref{sec:generality-correctness})

\subsection{Experimental Setup}

Unless otherwise stated, experiments use the default workload: Llama-3.1-8B~\cite{grattafiori2024llama3herdmodels}
with PPO~\cite{ouyang2022training} and vLLM~\cite{kwon2023vllm} for
rollout generation. Each comparison uses the same model, prompts, sampling parameters, maximum response length, and GPU budget at each step. Generation-stage latency includes rollout and CPU-side response processing.

\mypar{Testbeds}
We use three GPU clusters (\T\ref{tab:testbeds}) spanning accelerator vendors, interconnects, and scales.

\mypar{Baselines}
We compare \sys{} with OpenRLHF~\cite{hu2024openrlhf} and Verl~\cite{sheng2024hybridflow}
for synchronous and Laminar~\cite{sheng2025laminar} for asynchronous execution.
\sys{} runs on top of OpenRLHF, Verl, and Laminar.
For each baseline, we sweep parallelism and memory settings (\eg vLLM TP degree, Megatron TP/PP, ZeRO stage) and fix its fastest setting at startup per \mbox{GPU budget}.

For transition efficiency (\S\ref{sec:state-change}), we compare with UCP~\cite{lian2025universal}, MCP~\cite{megatron_dist_ckpt}, Gemini~\cite{wang2023gemini}, Tenplex~\cite{tenplex}, and Oobleck~\cite{jang2023oobleck}.
These baselines move the same parameter and optimizer state, and their setup time is counted.
We isolate DynaRL-style admission on held-out traces
(\S\ref{subsec:runtime-drift}) and DynaRL's per-component
migration in the coordination comparison (\S\ref{subsec:coordination-ablation}).

\mypar{Models and RL algorithms}
Beyond Llama-3.1-8B, we evaluate Qwen3-14B,
Qwen3-32B~\cite{yang2025qwen3technicalreport}, and
Llama-3.3-70B~\cite{grattafiori2024llama3herdmodels}.
Generality experiments replace PPO with ReMax~\cite{li2024remax} or
GRPO~\cite{shao2024deepseekmath}.
All end-to-end workloads use prompts from OpenRLHF \code{prompt-collection-v0.1}.
For each workload, the actor, reference, reward, and (when present) critic models have equal size.

\mypar{Metrics}
We measure end-to-end throughput (generated response tokens per RL step divided by step latency), step latency (including transitions), stage latency, and
transition cost. Transition cost is the time spent on setup (communicator creation) and on moving or resharding parameter and optimizer state
during one transition.
We also report reward, the PPO KL estimate, and the gap to the empirical optimum.
Unless stated otherwise, each latency and throughput value averages 50 steps after one warm-up step, with a max--min step-latency spread below 2\%.

\subsection{End-to-End Performance}
\label{sec:overall-performance}

\begin{figure}[t]
	\centering
	\begin{subfigure}[b]{1\linewidth}
		\centering
		\includegraphics[width=1\linewidth]{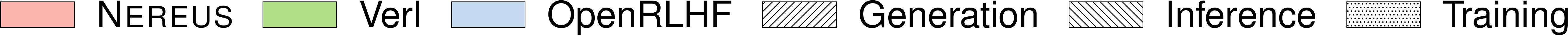}
	\end{subfigure}
	\begin{subfigure}[t]{0.49\linewidth}
		\centering
		\includegraphics[width=1\linewidth]{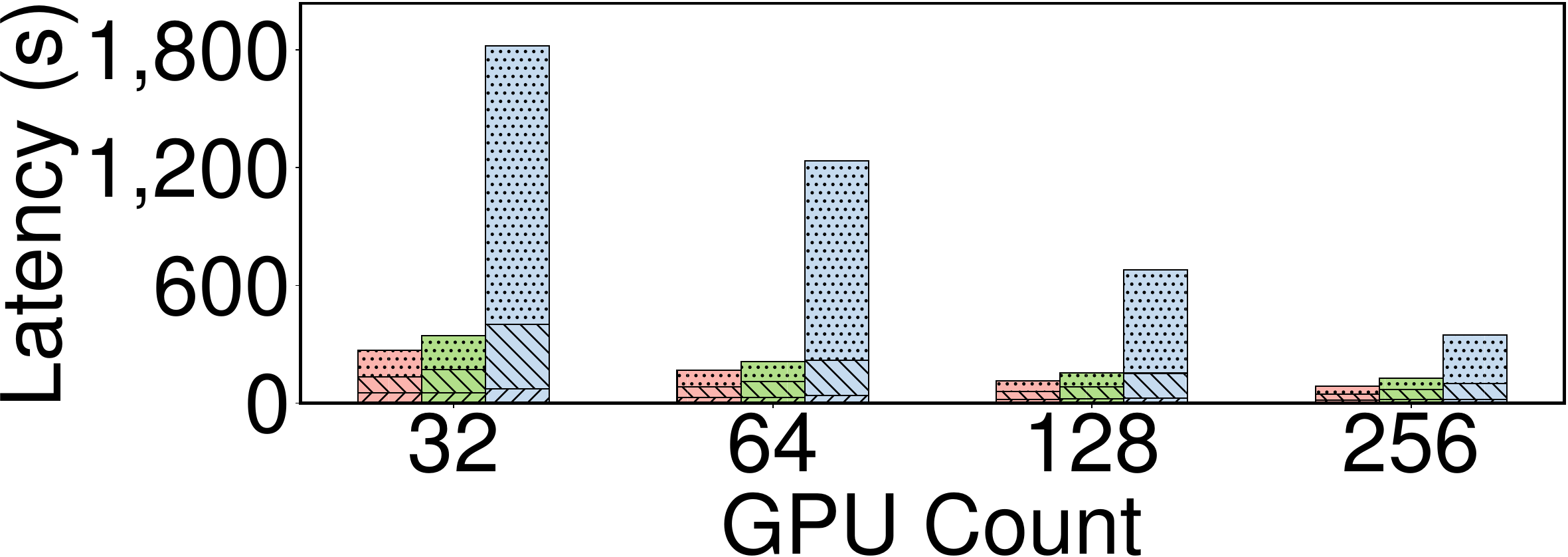}
		\caption{Step latency}
		\label{fig:leonardo-8b-latency}
	\end{subfigure}
	\hfill
	\begin{subfigure}[t]{0.49\linewidth}
		\centering
		\includegraphics[width=1\linewidth]{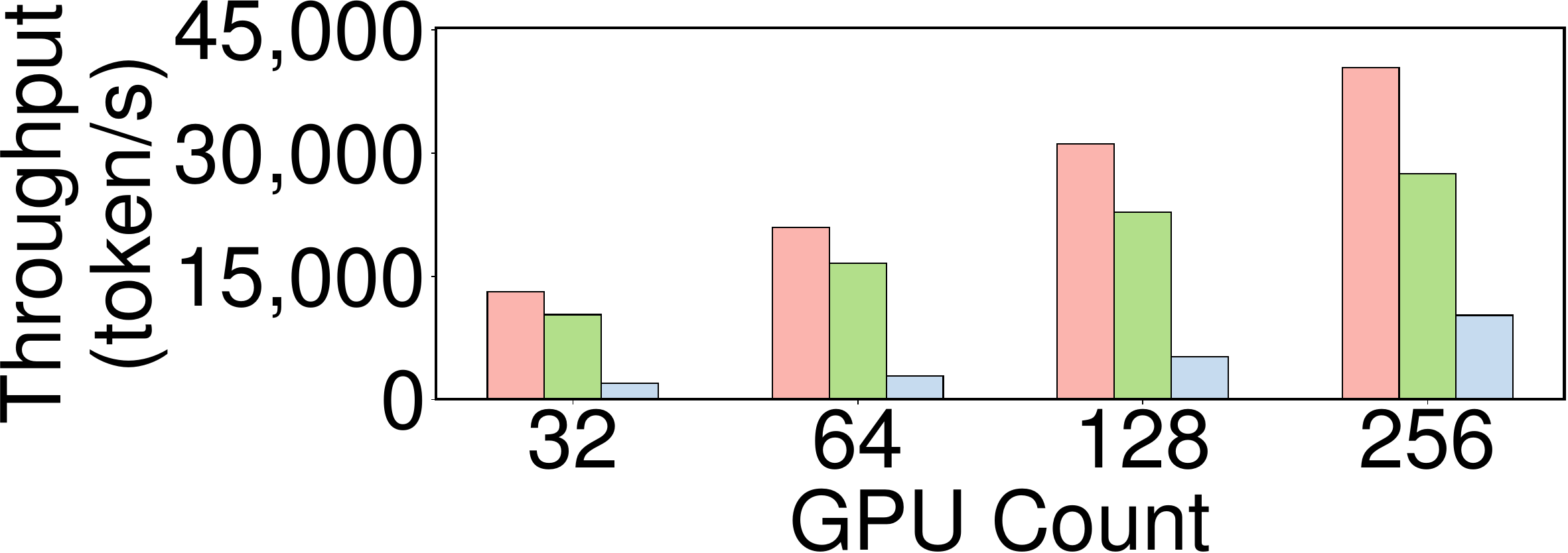}
		\caption{Throughput}
		\label{fig:leonardo-8b-throughput}
	\end{subfigure}
\caption{Llama-3.1-8B PPO performance at fixed GPU budgets on Cluster~\#2.}
	\label{fig:8b-performance-overall}
\end{figure}

{\sloppy We first compare complete-system performance against OpenRLHF and Verl.\footnote{We evaluate Verl only on NVIDIA GPUs because, at the time of writing, it did not support our AMD MI250X setup.}
At fixed GPU budgets, adaptation is primarily \emph{workload-driven}.
Within each run, \sys{} adapts to changing sequence lengths and stage bottlenecks, and we report the results across multiple fixed budgets.
\S\ref{subsec:runtime-drift}
then evaluates TP/PP adaptation and admission on traces built from real data, with the same runtime and transition engine.\par}

\subsubsection{Overall Performance}

Built on Verl, \sys{} outperforms both baselines on Cluster~\#2 (\F\ref{fig:8b-performance-overall}). It reduces step latency by up to 86.3\% relative to OpenRLHF and 31.9\% relative to Verl. Throughput increases by up to 7.27$\times$ and 1.47$\times$, respectively.
\sys{} reduces stage latency by up to 11.5\%, 41.6\%, and 29.6\% versus Verl for
generation, inference, and training, respectively, with the same trend on Cluster~\#3.
For the Llama-3.1-8B PPO comparisons, the speedup over \mbox{OpenRLHF} spans 2.14--7.27$\times$ (median 3.99$\times$) across all clusters and scales.
The 7.27$\times$ maximum occurs on Cluster~\#2 at 64 GPUs. On NVIDIA clusters, the speedup over Verl spans 1.10--1.47$\times$ (median 1.21$\times$).

\subsubsection{Scalability}

\begin{figure}[t]
	\centering
	\begin{minipage}{1\linewidth}
		\centering
		\begin{subfigure}[b]{1\linewidth}
			\centering
			\includegraphics[width=0.9\linewidth]{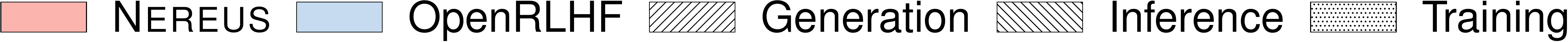}
		\end{subfigure}
		\begin{subfigure}[t]{0.49\linewidth}
			\centering
			\includegraphics[width=\linewidth]{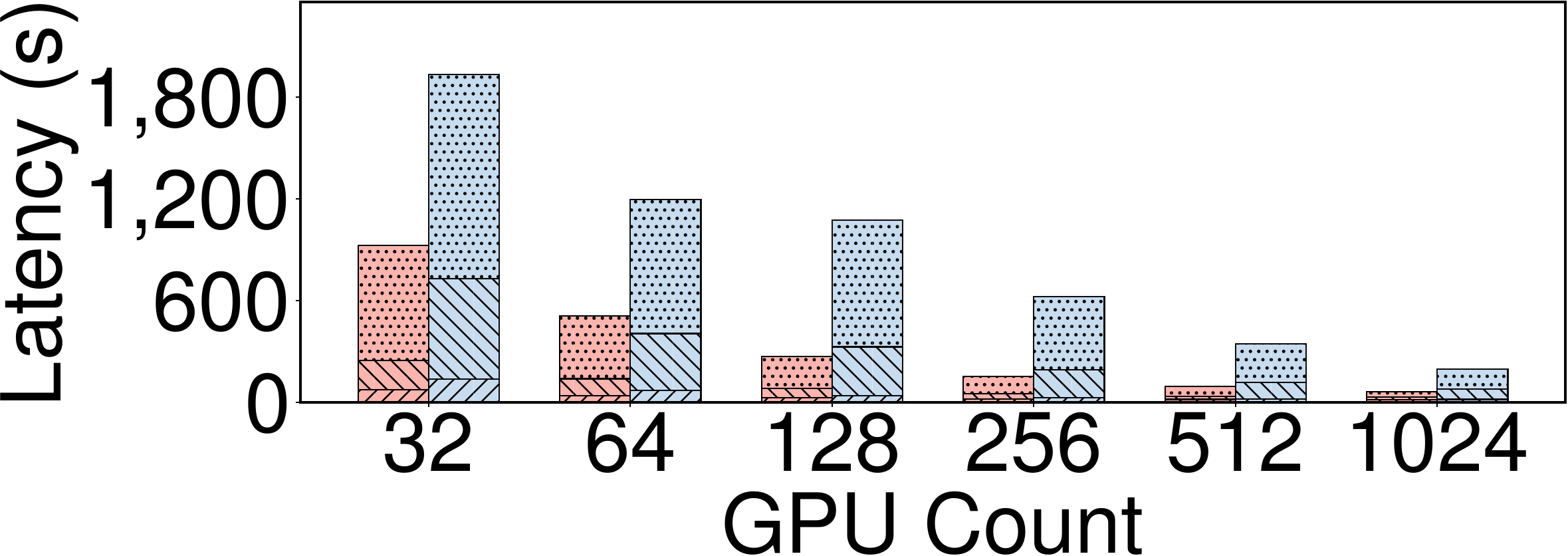}
			\caption{Step latency}
			\label{fig:lumi-time-scaling}
		\end{subfigure}
		\hfill
		\begin{subfigure}[t]{0.49\linewidth}
			\centering
			\includegraphics[width=\linewidth]{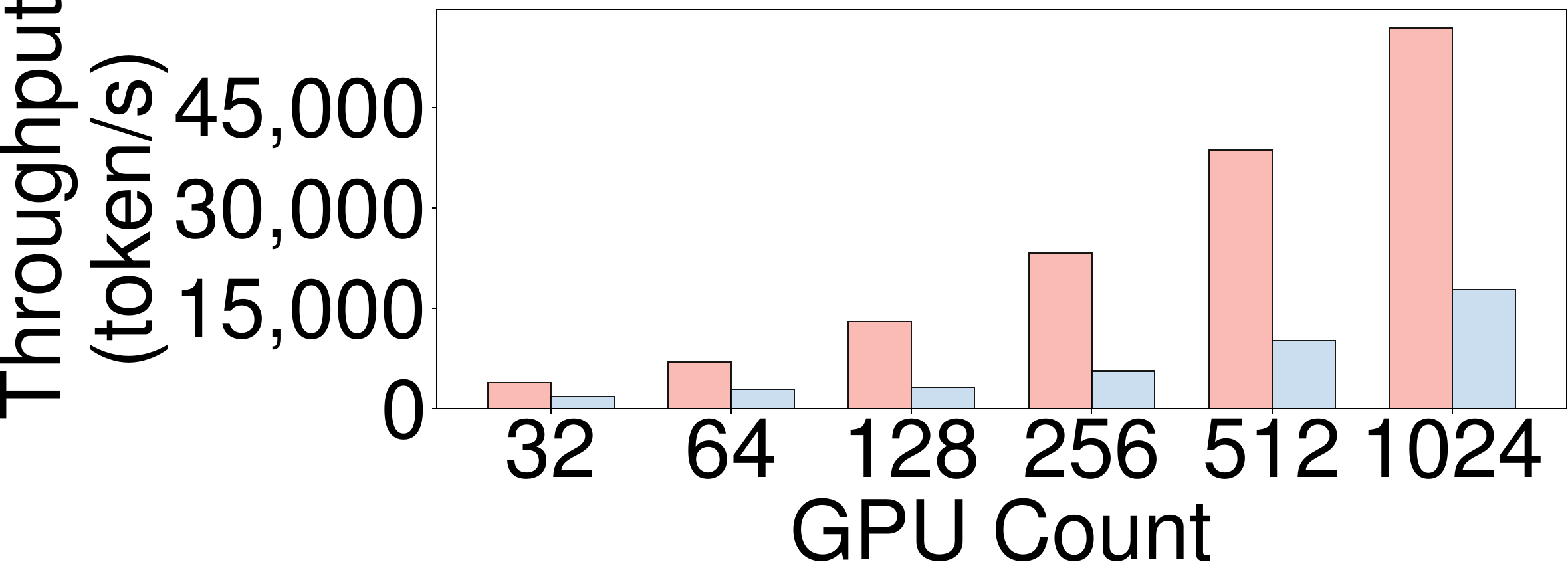}
			\caption{Throughput}
			\label{fig:lumi-throughput-scaling}
		\end{subfigure}
		\caption{Strong scalability on Cluster~\#1.}
		\label{fig:strong-scalability-merged}
	\end{minipage}
\end{figure}

\begin{figure}[t]
	\centering
	\includegraphics[width=1\linewidth]{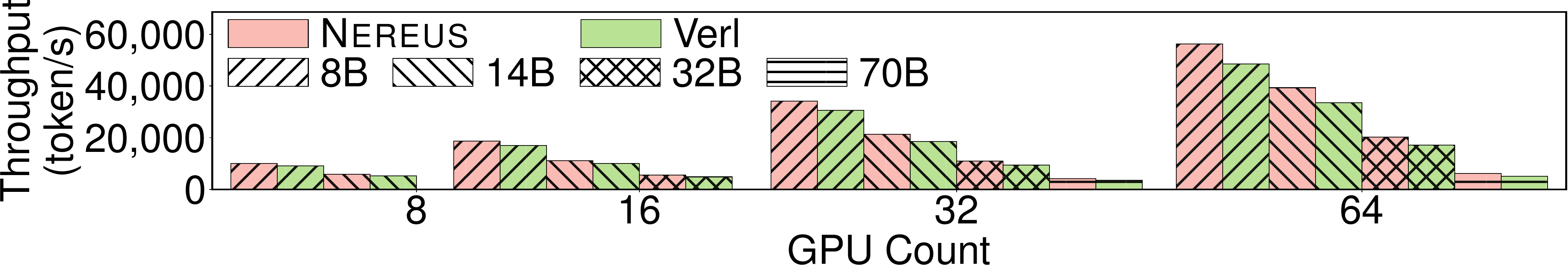}
\caption{PPO throughput across 8B--70B models and GPU budgets on Cluster~\#3.}
	\label{fig:weak-scalability-throughput}
\end{figure}

These end-to-end gains also persist at larger scales, up to 1{,}024 GPUs on Cluster~\#1.
From 32 to 1{,}024 GPUs, \sys{}'s step latency falls by 15$\times$, versus 10$\times$ for OpenRLHF. At 1{,}024 GPUs, its end-to-end speedup over OpenRLHF is 3.20$\times$ (\F\ref{fig:strong-scalability-merged}).
On Cluster~\#2, scaling from 32 to 256 GPUs reduces \sys{}'s step latency by 3.1$\times$, versus 2.7$\times$ for Verl (\F\ref{fig:leonardo-8b-latency}).

We further compare models from 8B to 70B across their feasible GPU budgets on Cluster~\#3.
\sys{} outperforms Verl for every model size, so the end-to-end gains persist beyond the 8B setting (\F\ref{fig:weak-scalability-throughput}).

\subsection{Cost-Aware Adaptation Policy}
\label{sec:cost-model}

We evaluate cost-model calibration and overhead, plan selection against empirical optima, and admission under drift.

\subsubsection{Online Cost-Model Calibration and Overhead}
\label{subsec:cost-model-efficacy}

\begin{figure}[t]
	\centering
	\begin{minipage}{1\linewidth}
		\centering
		\begin{subfigure}[b]{0.48\linewidth}
			\centering
			\includegraphics[width=\linewidth]{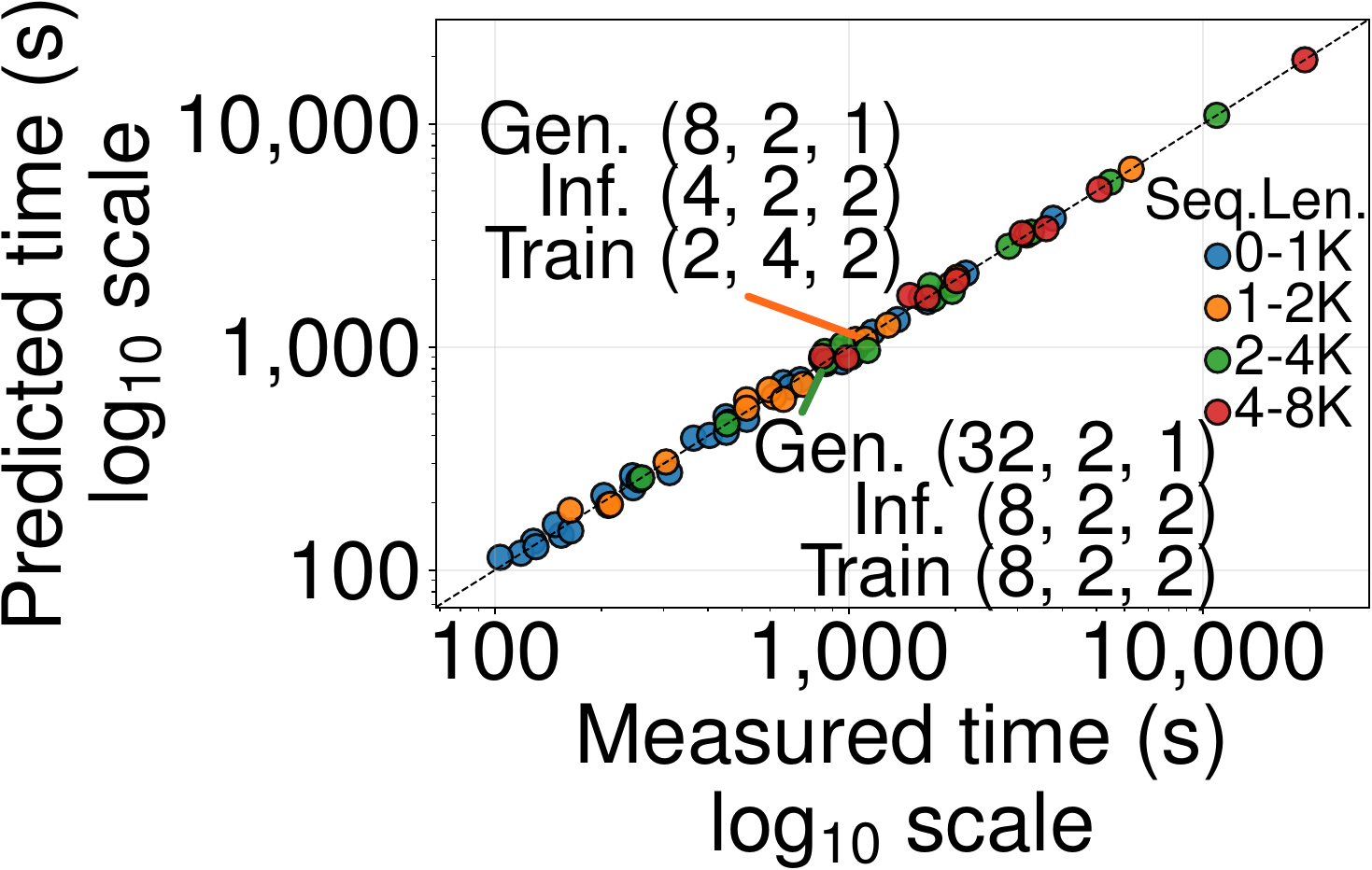}
			\caption{Calibration fit}
			\label{fig:cost-model-efficacy}
		\end{subfigure}
		\hfill
		\begin{subfigure}[b]{0.48\linewidth}
			\centering
			\includegraphics[width=\linewidth]{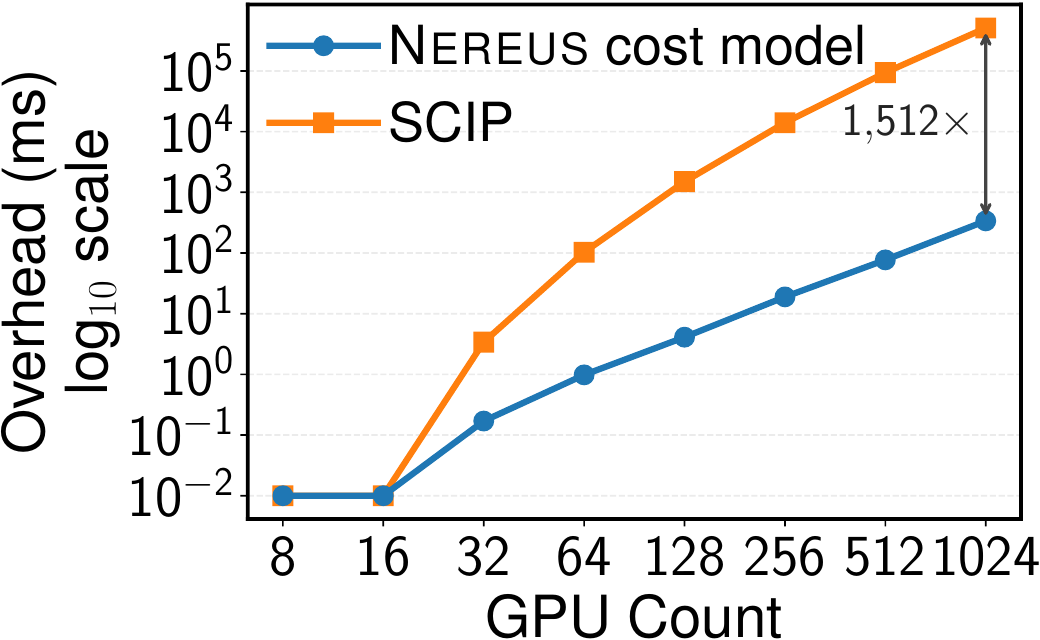}
			\caption{Decision overhead}
			\label{fig:cost-model-overhead}
		\end{subfigure}
\caption{Cost-model fit to measured training-stage latency (a) and target-plan selection time (b). Annotations in (a) list the (DP, TP, PP) tuples for two example global plans.}
		\label{fig:cost-model}
	\end{minipage}
\end{figure}

For the default workload on Cluster~\#2, we evaluate the online cost model's training-stage latency term on 72 measured
plans spanning 8--256 GPUs and sequence-length buckets of 0--1K, 1--2K, 2--4K,
and 4--8K tokens. This term is calibrated only on operation-level kernel and collective timings (\S\ref{subsec:cost}) from separate runs, so these plans are out-of-sample. The predictions have a 4.96\%
mean absolute percentage error and 14.61\% maximum error, with $R^2=0.9993$
(\F\ref{fig:cost-model-efficacy}).

We compare \sys{}'s decision overhead with a SCIP-based solver~\cite{hojny2025scip} for Eq.~(\ref{prob:control}).
\sys{} keeps the decision overhead low (\F\ref{fig:cost-model-overhead}), from 0.17\unit{ms} at 32 GPUs to 338\unit{ms} at 1{,}024 GPUs, whereas solver-based search grows from 3.4\unit{ms} to 511\unit{s}.
Peak replanning memory stays below 100\unit{MB}.

\subsubsection{Closeness to Empirical Optima}
\label{subsec:empirical-optima}

For the default workload on Cluster~\#2, we enumerate and execute 480 valid plans
across 18 settings (sequence lengths 1K, 2K, and 4K at GPU counts
8, 16, 32, 64, 128, and 256). The minimum measured step latency in each setting is
the empirical optimum within the controller's plan space. Each plan is
executed directly, without a transition. The 480 plans are also held out and do not overlap the 72 plans of \S\ref{subsec:cost-model-efficacy}. \sys{} exactly matches the empirical optimum in 61.1\%
of settings and stays within 5\% in all settings, with a 0\% median gap and a
95th-percentile (p95) gap of 4.1\%.

\subsubsection{Decision Quality Under Dynamic Runtime Drift}
\label{subsec:runtime-drift}

\begin{figure}[t]
	\centering
	\includegraphics[width=0.95\linewidth]{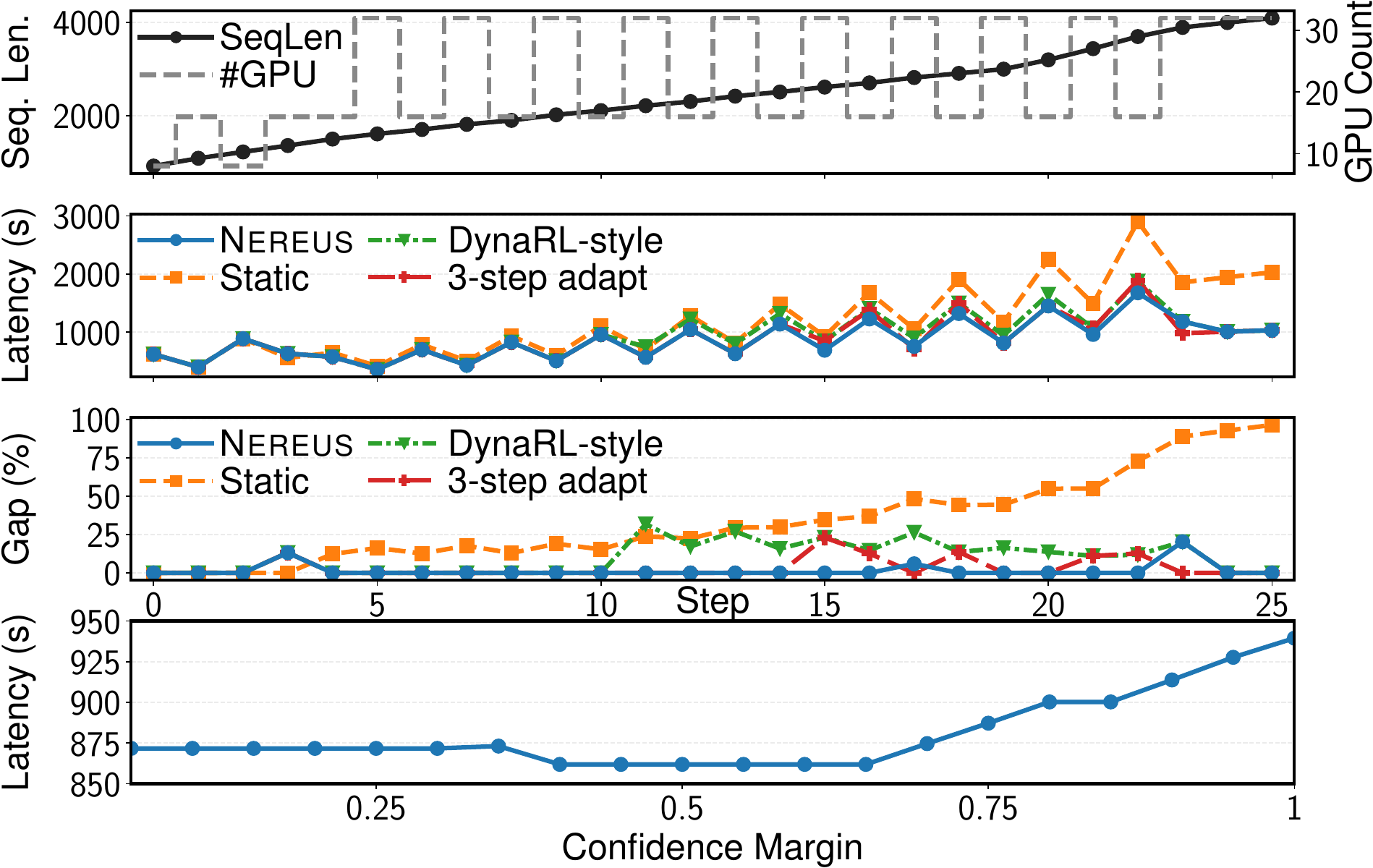}
\caption{Controller policies under held-out sequence-length and GPU-supply drift: step latency, gap to the best plan at each step, and sensitivity to the confidence margin.}
	\label{fig:dynamic-trace-controller}
\end{figure}

We run four policies on Cluster~\#2 over held-out sequence-length and GPU traces, with the same runtime, planner, and
\abs{} primitives. The \code{static} policy keeps its initial TP/PP layout, and \code{3-step adapt} always switches to the planner's target every three steps.
All policies scale DP with the available GPUs. Under DynaRL-style admission, any transition with $\Delta L>0$ is admitted,
while \sys{} also requires the savings to repay the transition cost. We repeat the evaluation on two more held-out
traces with the same settings and plans.

All traces are built from real data (\Fs\ref{fig:gpu-ava} and~\ref{fig:sequence-length}). In the first trace (\F\ref{fig:dynamic-trace-controller}), average sequence length grows
from 1{,}147.5 tokens (steps 0--3) to 3{,}722 tokens (steps 20--25). The p95
length grows from 1{,}339 to 4{,}072 tokens, while available GPUs vary from 8 to 32.
Under this drift, \sys{} admits two TP/PP transitions, at steps 3 and 23, together with DP~scaling.

\sys{} achieves the lowest average step latency (861.9\unit{s}, versus 1{,}191.6\unit{s} for \code{static}), reducing latency by 27.7\% over the initial fixed TP/PP layout, 8.3\% over DynaRL-style admission, and 2.5\% over \code{3-step adapt}.
In the confidence-margin sweep (\F\ref{fig:dynamic-trace-controller}, bottom), latency is lowest at $\gamma=0.4$--$0.65$ (861.9\unit{s}, 2 transitions). At $\gamma=1$, latency reaches 939.5\unit{s} with 14 transitions, matching DynaRL-style admission on the first trace, as the extra transitions cost more than they save.

Across three held-out traces, each run five times on hardware, \sys{} ($\gamma=0.5$, fixed before these experiments) averages
858.7\unit{s}/step (5.2\unit{s} standard deviation across traces), versus 873.8\unit{s} (12.6\unit{s}) for
\code{3-step adapt} and 928.3\unit{s} (19.7\unit{s}) for DynaRL-style admission. All runs use a 32-GPU allocation, within which \sys{} releases and reacquires GPUs as the trace changes. The five runs of each trace differ by less than 1\%. \sys{} makes
two TP/PP transitions per trace, each taking 6.5--16\unit{s}, and no admitted plan ran out of memory.

\subsection{\abss{} and Transition Orchestration}
\label{sec:state-change}

We measure the cost of the \abs{} boundary, primitives, and GPU-direct transport (\S\ref{subsec:resource-scale}--\S\ref{subsec:para-overhead}), DAG planning time (\S\ref{subsec:transition-overhead}), and cross-model-stage coordination (\S\ref{subsec:coordination-ablation}).

\subsubsection{Resource-Scaling Cost}
\label{subsec:resource-scale}

\begin{table}[t]
	\centering
	\small
\caption{Cost (s) of doubling the GPUs for the 8B actor/critic workload on Cluster~\#3. $^\dagger$Also persists state.}
	\label{tab:reconf_overhead_cluster3}
	\begin{tabular}{lcccc}
		\toprule
		\textbf{System} & \textbf{4$\to$8} & \textbf{8$\to$16} & \textbf{16$\to$32} & \textbf{32$\to$64} \\
		\midrule
		UCP$^\dagger$ & 695.94 & 798.94 & 836.74 & 981.21 \\
		Tenplex & 39.72 & 51.68 & 66.43 & 89.92 \\
		Gemini$^\dagger$ & 26.56 & 37.12 & 56.83 & 87.25 \\
		Oobleck & 10.54 & 21.48 & 39.06 & 83.72 \\
		\sys{} & \textbf{2.45} & \textbf{5.58} & \textbf{6.52} & \textbf{8.48} \\
		\bottomrule
	\end{tabular}
\end{table}

We measure \code{Extend} cost for the 8B actor/critic workload on Cluster~\#3.
\code{Extend} takes 2.45--8.48\unit{s} from 4$\to$8 through 32$\to$64 GPUs (\T\ref{tab:reconf_overhead_cluster3}).
\sys{} is 3.8--9.9$\times$ faster than Oobleck, 6.7--10.8$\times$ faster than Gemini, 9.3--16.2$\times$ faster than Tenplex, and 115.7--284.1$\times$ faster than UCP (Gemini and UCP also persist state).

Transition cost remains a small fraction of run time at full cluster scale.
Six TP/PP transitions in a 1{,}000-step PPO run reaching 1{,}024 GPUs on Cluster~\#1 consume 49.5\unit{s} of 62{,}353\unit{s} (0.079\%).
The largest one, which also doubles the GPUs from 512 to 1{,}024, takes 31.55\unit{s}, versus 1{,}629\unit{s} for UCP. \code{Split} and \code{Extend} dominate it, and \code{Merge} and \code{Destroy} cost little.

\subsubsection{Parallelism-Resharding Cost}
\label{subsec:para-overhead}

\begin{figure}[t]
	\centering
	\begin{minipage}{1\linewidth}
		\centering
		\begin{subfigure}[b]{1\linewidth}
			\centering
			\includegraphics[width=1\linewidth]{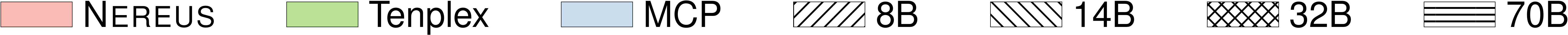}
		\end{subfigure}
		\begin{subfigure}[b]{1\linewidth}
			\centering
			\includegraphics[width=1\linewidth]{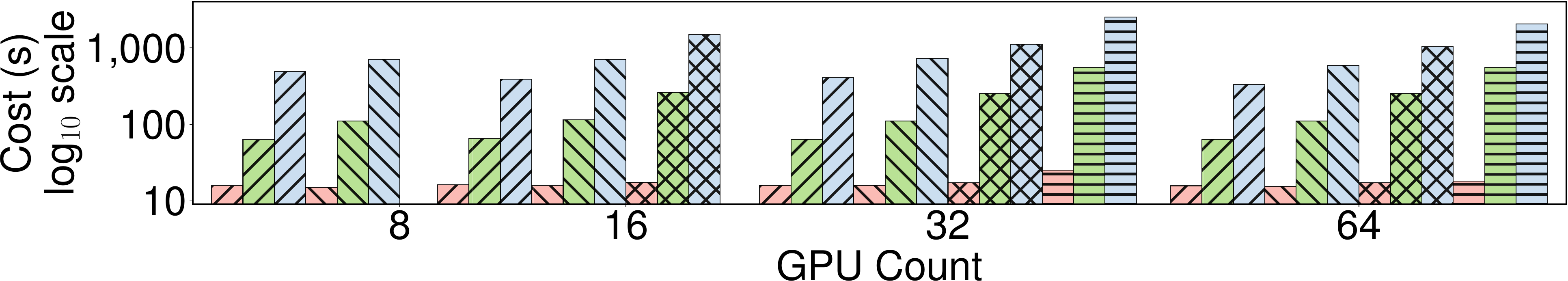}
		\end{subfigure}
		\caption{Parallelism-resharding cost on Cluster~\#3. MCP also persists state. Missing bars indicate OOM in all systems.}
		\label{fig:starting-time}
	\end{minipage}
\end{figure}

We measure TP/PP resharding using \code{Split} and \code{Merge} for models from 8B to 70B parameters.
We compare against MCP and Tenplex.
MCP (checkpoint-based) and Tenplex (shard-level) are the closest coarse- and fine-grained alternatives to \abs{} resharding.
Gemini and Oobleck appear only in the resource-scaling comparison
(\S\ref{subsec:resource-scale}). Gemini targets checkpoint-based recovery,
while Oobleck reconfigures pipelines under node changes and uses FSDP within
pipeline stages~\cite{wang2023gemini,jang2023oobleck}.
For a 70B model on 64 GPUs, \sys{} reduces transition cost by 99.1\% relative to MCP and by 96.7\% relative to Tenplex (\F\ref{fig:starting-time}).
For the 8B model, each bar averages \code{Split}/\code{Merge} transitions over the same feasible TP/PP layouts at each GPU count. Resharding takes a stable 15.7--16.1\unit{s} in \sys{} across 8--64 GPUs, versus 62.7--65.0\unit{s} for Tenplex's shard-level resharding and 329.9--486.5\unit{s} for MCP.

\subsubsection{Transition Planning}
\label{subsec:transition-overhead}

\begin{figure}[t]
	\centering
	\begin{minipage}{1\linewidth}
		\centering
		\begin{subfigure}[b]{0.48\linewidth}
			\centering
			\includegraphics[width=\linewidth]{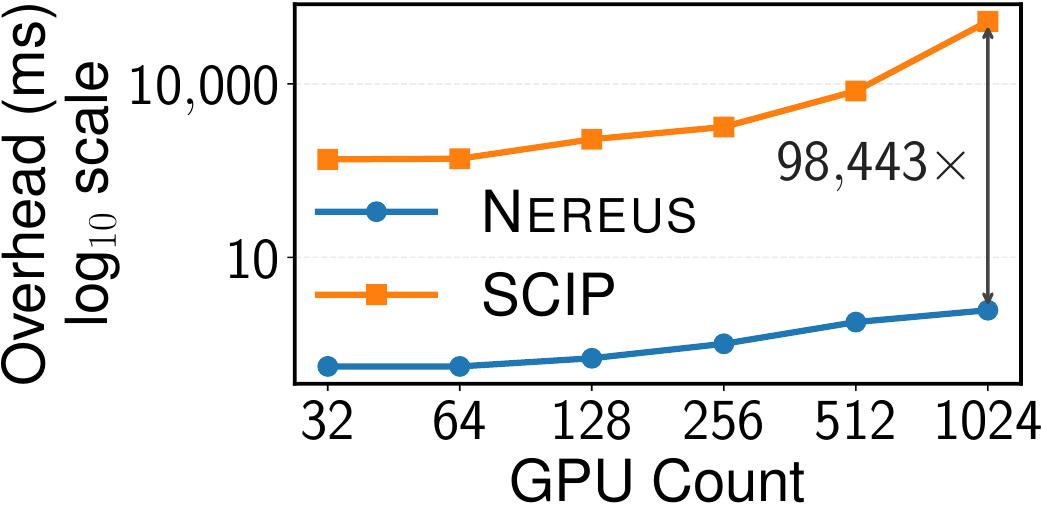}
			\caption{Planning overhead}
			\label{fig:transition-planning-time}
		\end{subfigure}
		\hfill
		\begin{subfigure}[b]{0.48\linewidth}
			\centering
			\includegraphics[width=\linewidth]{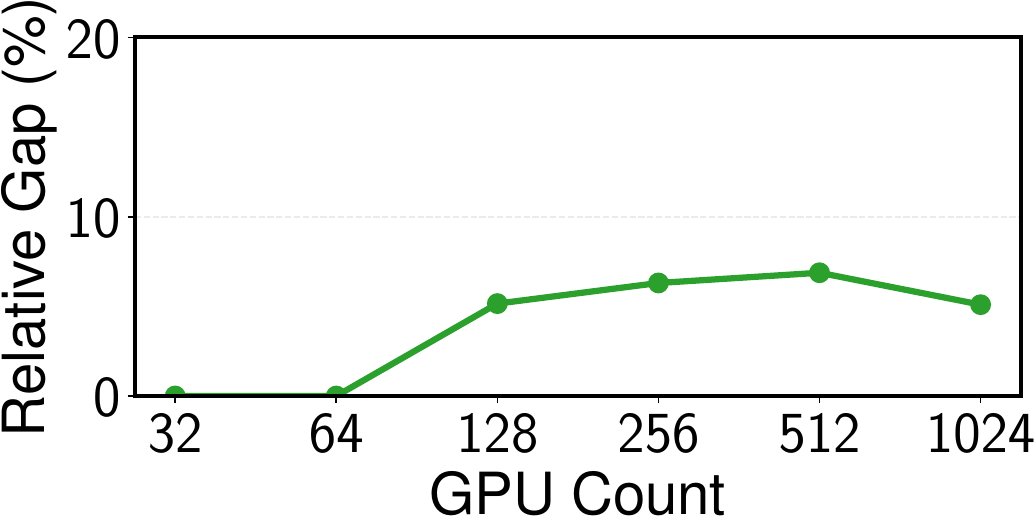}
			\caption{Gap to SCIP}
			\label{fig:transition-planning-gap}
		\end{subfigure}
		\caption{Transition planning across 32--1{,}024 GPUs.}
		\label{fig:transition-planning}
	\end{minipage}
\end{figure}

We compare transition-DAG planning with SCIP~\cite{hojny2025scip}, which solves the scheduling problem of \S\ref{subsec:dag-construction}. \sys{}'s greedy heuristic
produces similar schedules in less planning time (\F\ref{fig:transition-planning}).
At 1{,}024 GPUs, \sys{} takes 1.22\unit{ms} versus 120.10\unit{s} for SCIP. Planning is 98{,}443\(\times\) faster at this scale, where
SCIP's planning time exceeds \sys{}'s step latency (\F\ref{fig:lumi-time-scaling}). The relative gap to SCIP in transition completion time stays within 6.9\% at every scale (5.1\% at 1{,}024 GPUs).
This gap measures schedule quality within the same formulation, whereas \S\ref{subsec:empirical-optima}
reports target-plan quality against the empirical optimum.

\subsubsection{Cross-Model-Stage Coordination}
\label{subsec:coordination-ablation}

\begin{table}[t]
	\centering
	\footnotesize
	\setlength{\tabcolsep}{3.5pt}
\caption{Coordination under three transition types: success rate (\%), successful-trial time (s), and added resource-dependency edges. DynaRL-style uses per-component migration with independent local DAGs. SM-CS, CM-SS, and CM-CS denote same-model cross-stage, cross-model same-stage, and cross-model cross-stage overlap.}
	\label{tab:coordination-ablation}
	\begin{tabular}{lcccccc}
		\toprule
		\multirow{2}{*}{\textbf{Type}} & \textbf{Overlap} & \multicolumn{2}{c}{\textbf{DynaRL-style}} & \multicolumn{2}{c}{\textbf{\sys{}}} & \multirow{2}{*}{\textbf{Edges}} \\
		\cmidrule(lr){3-4}\cmidrule(lr){5-6}
		 & \textbf{GPUs} & \textbf{Succ.} & \textbf{Time} & \textbf{Succ.} & \textbf{Time} & \\
		\midrule
		SM-CS & 32 & 62 & 8.48 & 100 & 9.80 & 1 \\
		CM-SS & 48 & 56 & 8.93 & 100 & 10.80 & 2 \\
		CM-CS & 48 & 34 & 10.44 & 100 & 12.60 & 2 \\
		\bottomrule
	\end{tabular}
\end{table}

{\sloppy For the three transition types of \T\ref{tab:coordination-ablation}, we compare \sys{} on Cluster~\#3 with DynaRL's per-component migration, which runs the local DAGs independently.
We run 100 trials per system and type, with randomized launch timing and GPU assignment.
A trial succeeds only if all local DAGs finish within 20\unit{s} and reach the target
\abs{} collections without shared-GPU conflicts. \sys{} inserts 1--2 resource-dependency
edges and succeeds in all trials. DynaRL-style succeeds in
34--62\%, and each failure is a shared-GPU deadlock that a longer timeout cannot resolve. Counting failures as 20\unit{s} yields \mbox{12.86--16.75\unit{s}} for
DynaRL-style versus 9.80--12.60\unit{s} for \sys{}.\par}

\subsection{Generality and Training Behavior}
\label{sec:generality-correctness}

We assess performance beyond synchronous PPO and compare the observed training behavior of
\sys{} and Verl over equal RL-step counts.

\subsubsection{Generality Across Algorithms and Execution Modes}
\label{sec:grporemax}

\begin{figure}[t]
	\centering
	\begin{subfigure}[b]{0.32\linewidth}
		\centering
		\includegraphics[width=\linewidth]{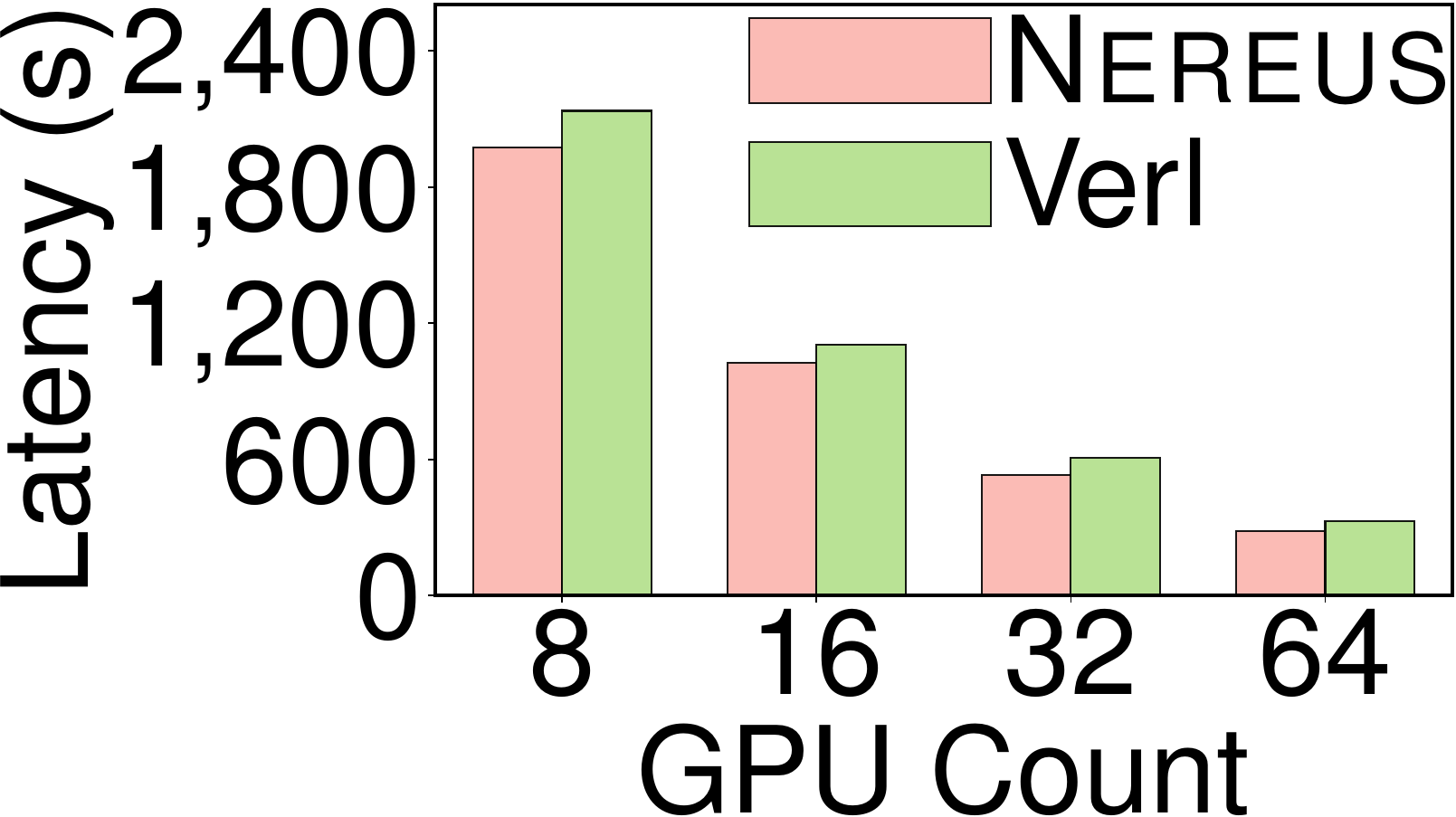}
		\caption{ReMax}
		\label{fig:triton_8b_remax_8_step}
	\end{subfigure}
	\begin{subfigure}[b]{0.32\linewidth}
		\centering
		\includegraphics[width=\linewidth]{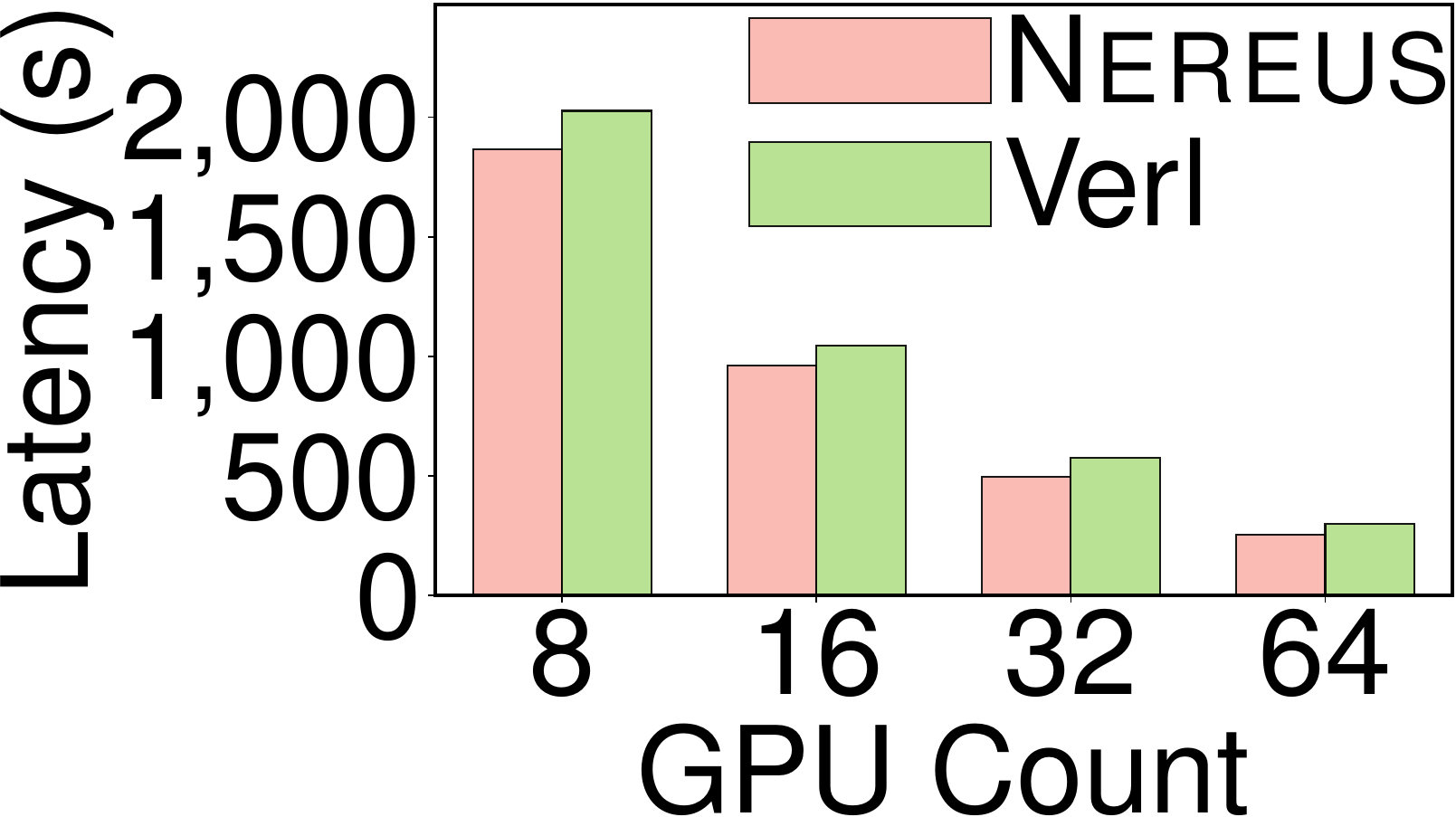}
		\caption{GRPO}
		\label{fig:triton_8b_grpo_8_step}
	\end{subfigure}
	\begin{subfigure}[b]{0.32\linewidth}
		\centering
		\includegraphics[width=\linewidth]{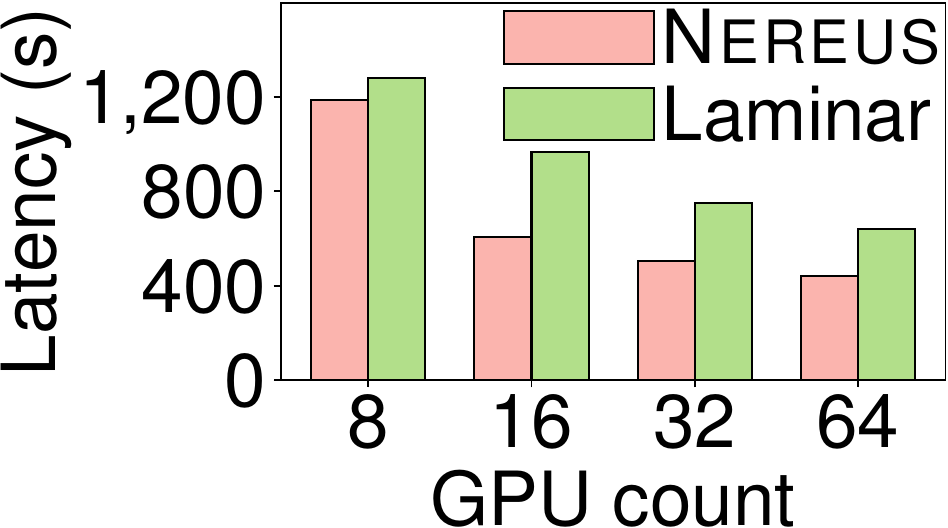}
		\caption{Async. RL}
		\label{fig:triton_8b_async_step}
	\end{subfigure}
	\caption{Step latency on Cluster~\#3 for ReMax and GRPO (eight samples per prompt) and asynchronous RL.}
	\label{fig:triton-grpo-remax}
\end{figure}

\begin{figure}[t]
	\centering
	\includegraphics[width=\linewidth]{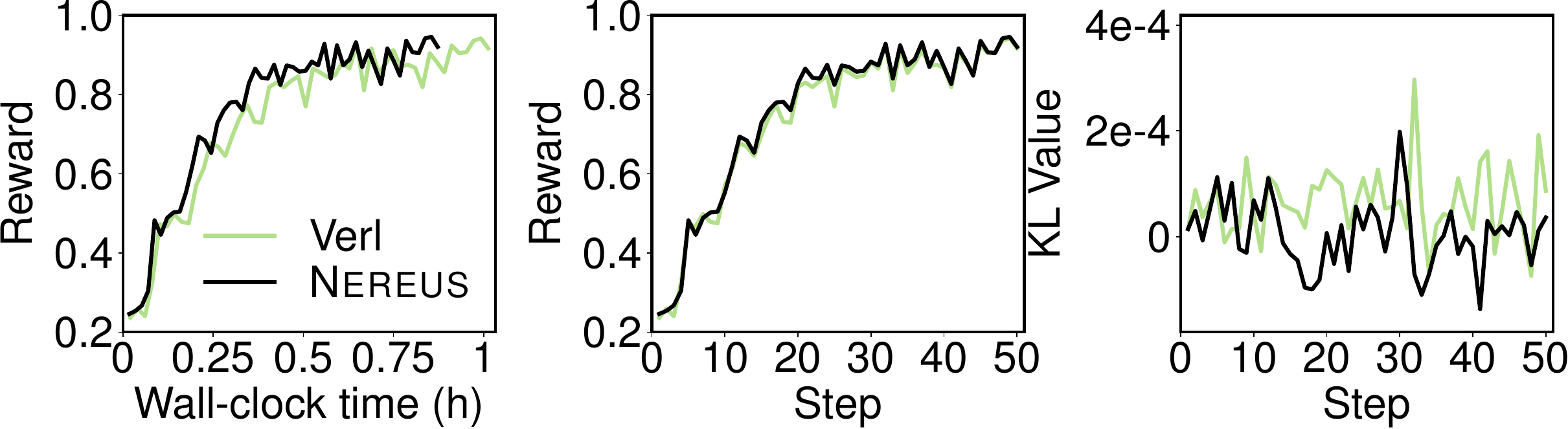}
	\caption{Training behavior over the first 50 steps: reward against wall-clock time (left), reward and PPO KL estimate against step (middle, right).}
	\label{fig:convergence}
\end{figure}

On Cluster~\#3, with ReMax and GRPO, \sys{} outperforms Verl at every
GPU count (\F\ref{fig:triton-grpo-remax}), reducing latency by up to 13.3\% for ReMax
and 15.8\% for GRPO.

We also evaluate \sys{} in a fully asynchronous RL setting, where it achieves up to 37.3\% lower step latency than Laminar~\cite{sheng2025laminar} (\F\ref{fig:triton_8b_async_step}).
Here, step boundaries are weight synchronizations between training and rollout,
with adaptation subject to the safe-boundary conditions in \S\ref{subsec:monitor}.

\subsubsection{Training Behavior Under Adaptation}
\label{sec:correctness}

Over the first 50 steps, \sys{} tracks Verl's reward and PPO KL estimate (the mean log-probability drop of sampled tokens from the rollout policy to the updated policy,
\F\ref{fig:convergence}). Both systems reach reward 0.90 at step 32. \sys{} reaches
0.94 at step 48, and Verl at step 49. At step 50, \sys{}'s reward is 0.9199 versus 0.9160 for Verl,
and the mean absolute PPO KL estimate over the 50 steps is $4.9\times10^{-5}$
versus $7.4\times10^{-5}$.

In wall-clock terms, \sys{} reaches reward 0.90 after 0.56\unit{h} versus
0.65\unit{h} for Verl, and 0.94 after 0.84\unit{h} versus 0.99\unit{h}. The
same 50 steps, including four transitions, complete in 3{,}141\unit{s} instead of
3{,}646\unit{s}, a 13.9\% wall-clock reduction relative to
Verl, with closely matched reward and KL curves.

\section{Related Work}
\label{sec:related-work}

\myparr{RL post-training systems} optimize synchronous and asynchronous execution.
Synchronous frameworks (\eg Verl~\cite{sheng2024hybridflow}) improve
flexibility and efficiency~\cite{xiao2025flexrlhf,hu2024openrlhf,nemo-rl,lei2024puzzle}, while
asynchronous frameworks (\eg Laminar~\cite{sheng2025laminar}) run generation,
inference, and training concurrently to handle varying trajectory lengths~\cite{wu2025llamarl,slime,wang2025roll,fu2025areal,zhou2025aprilactivepartialrollouts,han2025asyncflowasynchronousstreamingrl,gao2025rollartscalingagenticrl}.
ReaL~\cite{mei2025real} and ROLL~\cite{wang2025roll} optimize initial plans or placements.
StreamRL~\cite{zhong2025streamrl} adds generation replicas online but restarts to reconfigure trainers.
DynaRL~\cite{wang2026dynarl} reallocates GPUs within a step and a fixed pool, based on sustained underutilization and predicted global throughput gains.
\sys{} instead adapts across steps and pool changes, with payback-based admission, dependency-aligned \abss{}, and
safe concurrent transition orchestration.

\myparr{Cluster schedulers}~\cite{qiao2021pollux,jayaram2023sia,gu2023elasticflow} resize or reassign resources across jobs but do not coordinate plan transitions within a running RL post-training job.
\sys{} applies the payback principle (\S\ref{subsec:controller}) inside one coupled job.

\myparr{State-management systems} focus on selective state manipulation or migration for a single model under changing parallelism and resources~\cite{li2023easyscale,tenplex,jang2023oobleck,thorpe2023bamboo}.
They establish the value of reusing GPU-resident state but do not coordinate
multi-model, multi-stage adaptation. Checkpoint systems
manage state coarsely for fault tolerance~\cite{wang2023gemini} and incur restart pauses during plan transitions~\cite{zeng2026gcr,wan2025bytecheckpoint,pytorch_dist_ckpt,megatron_dist_ckpt,lian2025universal}. \sys{} reuses GPU-resident state per model-stage replica and coordinates changes across model-stages that share GPUs.

\myparr{Transfer engines} such as fabric-lib~\cite{licker2025rdma} provide RDMA
point-to-point paths for KV cache transfer, MoE dispatch, and RL weight
synchronization. They complement \sys{}, which decides which state moves and in what order.

\section{Conclusion}

Plan adaptation becomes tractable when the state boundary matches the job's dependencies. \sys{} decides when to adapt, what state to reuse, and how to transition, through a cost-aware adaptation policy, a dependency-aligned state abstraction, and safe concurrent transition
orchestration. Its selected plans stay within 5\% of the empirical
optimum in all 18 measured settings. In a real-data trace, it lowers average step latency by
27.7\% versus the initial fixed TP/PP layout. Across three held-out traces, it averages
858.7\unit{s}/step, versus 928.3\unit{s} for DynaRL-style admission. Six transitions in a 1{,}000-step run reaching 1{,}024 GPUs take
0.079\% of run time. Scaling is 3.8--16.2$\times$ faster than Oobleck and Tenplex, and coordinated transitions succeed in all overlap trials, versus \mbox{34--62\%} for DynaRL's per-component migration. \sys{} improves 8B PPO throughput by 2.14--7.27$\times$ over OpenRLHF
and by 1.10--1.47$\times$ over Verl across the evaluated clusters. The gains extend to ReMax, GRPO, and asynchronous RL.
The model-stage boundary and modular runtime also provide a path to extend adaptation to
context/expert parallelism, autoscaling external tool services, and additional RL frameworks.

\section*{Acknowledgments}

This work was supported by the Research Council of Finland
(Grant Nos. 362729 and 358877), Business Finland (Grant No. 169/31/2024),
and the Finnish Ministry of Education and Culture's Doctoral Education Pilot
through the Finnish Doctoral Program Network in Artificial Intelligence
(AI-DOC, Decision No. VN/3137/2024-OKM-6). The authors acknowledge
computing resources of EuroHPC JU projects
(EHPC-REG-2025R02-367, EHPC-DEV-2024D09-039 and EHPC-DEV-2025D10-012),
Aalto Science-IT project and CSC -- IT Center for Science, Finland.

\bibliographystyle{ACM-Reference-Format}
\Urlmuskip=0mu plus 1mu\relax
\providecommand{\showURL}[1]{\unskip}
\bibliography{sample-base}

\end{document}